\pdfoutput=1

\documentclass[sigconf,nonacm,10pt]{acmart}
\AtBeginDocument{%
  \providecommand\BibTeX{{%
    \normalfont B\kern-0.5em{\scshape i\kern-0.25em b}\kern-0.8em\TeX}}}

\setcopyright{acmcopyright}
\copyrightyear{2023}
\acmYear{2023}
\acmDOI{XXXXXXX.XXXXXXX}

\usepackage{multirow}
\usepackage[hypcap=true]{caption}
\usepackage{subfigure}
\usepackage{enumitem}
\usepackage{multirow}
\usepackage{caption}
\usepackage{tabularx}
\usepackage{amsmath}
\usepackage{afterpage}

\usepackage[linesnumbered,ruled,vlined]{algorithm2e}

\begin{document}
\title{\huge BRIEDGE: EEG-Adaptive Edge AI for Multi-Brain to Multi-Robot Interaction}

\author{Jinhui Ouyang}
\email{oldyoung@hnu.edu.cn}
\affiliation{%
  \institution{Hunan University}
  \city{Changsha}
  \state{Hunan}
  \country{China}
}

\author{Mingzhu Wu}
\email{wumz@hnu.edu.cn}
\affiliation{%
  \institution{Hunan University}
  \city{Changsha}
  \state{Hunan}
  \country{China}}

\author{Xinglin Li}
\email{lixinglin@hnu.edu.cn}
\affiliation{%
  \institution{Hunan University}
  \city{Changsha}
  \state{Hunan}
  \country{China}}

\author{Hanhui Deng}
\email{denghanhui@hnu.edu.cn}
\affiliation{%
  \institution{Hunan University}
  \city{Changsha}
  \state{Hunan}
  \country{China}}
  
\author{Di Wu}
\email{dwu@hnu.edu.cn}
\affiliation{%
  \institution{Hunan University}
  \city{Changsha}
  \state{Hunan}
  \country{China}}

\begin{abstract}
Recent advances in EEG-based BCI technologies have revealed the potential of brain-to-robot collaboration through the integration of sensing, computing, communication, and control. In this paper, we present BRIEDGE as an end-to-end system for multi-brain to multi-robot interaction through an EEG-adaptive neural network and an encoding-decoding communication framework as illustrated in Figure~\ref{fig_abstract}. As depicted, the edge mobile server or edge portable server will collect EEG data from the users and utilize the EEG-adaptive neural network to identify the users' intentions. \textcolor{black}{The encoding-decoding communication framework then encodes the EEG-based semantic information and decodes it into commands in the process of data transmission.} 
To better extract the joint features of heterogeneous EEG data as well as enhance classification accuracy, BRIEDGE introduces an informer-based ProbSparse self-attention mechanism. Meanwhile, parallel and secure transmissions for multi-user multi-task scenarios under physical channels are addressed by dynamic autoencoder and autodecoder communications. From mobile computing and edge AI perspectives, model compression schemes composed of pruning, weight sharing, and quantization are also used to deploy lightweight EEG-adaptive models running on both transmitter and receiver sides. Based on the effectiveness of these components, a code map representing various commands enables multiple users to control multiple intelligent agents concurrently. Our experiments in comparison with state-of-the-art works show that BRIEDGE achieves the best classification accuracy of heterogeneous EEG data, and more stable performance under noisy environments.

\end{abstract}



\begin{teaserfigure}
  \includegraphics[width=\columnwidth]{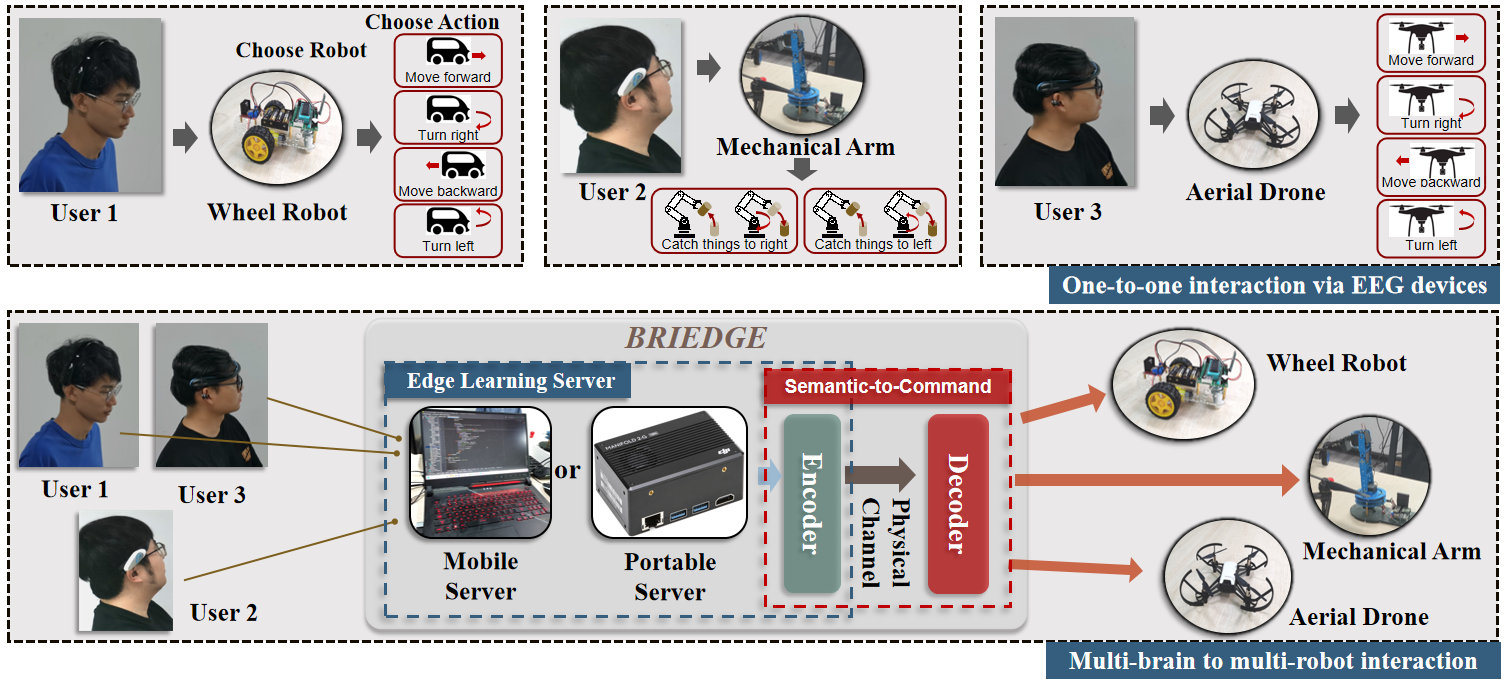}
  \caption{\textcolor{black}{BRIEDGE system in use. Multiple users wearing various EEG devices can control multiple robots to perform specific actions in the system by only thinking in their minds. We use the participants' photos with their content in this paper.}}
  \label{fig_abstract}
\end{teaserfigure}
\maketitle

\renewcommand{\shorttitle}{BRIEDGE: EEG-Adaptive Edge AI for Multi-Brain to Multi-Robot Interaction}

\section{Introduction}
The development of the brain-computer interface (BCI) is of immense significance as it establishes a vital bridge between the human brain and external equipment. In recent years, substantial advancements have been made in the design of BCI, in both academy and industry, through the emerging confluence of machine learning (ML) and human-computer interaction (HCI)~\cite{Neuralink,CaiMobiCom2018,HouMobiCom2019,kirill2019imwut}. Electroencephalography (EEG) signals are often collected and translated into different user intentions by a head-worn device in BCI systems for various applications, such as mind-controlled robot operations~\cite{Leeb2015} and web explorations~\cite{Moshfeghi2018WWW}. Therefore, BCI has the great potential to emerge as a prominent mainstream interaction method in the future, facilitating seamless connections among humans, machines and things.

Meanwhile, it has been witnessed by significant progress of deep learning in recent years. The success of Convolutional Neural Network (CNN) in computer vision area has inspired several works to enhance the EEG recognition accuracy with its capability in extracting the spatial structure feature of data~\cite{SakhaviTNNLS2018,NahmiasKDD2020}. Due to the strong temporal dependency of EEG signal, Recurrent Neural Network (RNN) model was also utilized to perform its advantage in handling sequential data~\cite{TaoTAC2020,ZhangTIST2020,zhang2018mindid}. Similarly, the Transformer model emerging in recent years is also regarded as an effective architecture when dealing with sequential data, which is worth noting in EEG recognition research as well~\cite{WuUbiComp2022}. 
However, existing works either neglect the subtle temporal dependency within EEG signals, or do not consider the spatial dependency across channels. Time dependency is one of the important features of EEG data, which has a great influence on EEG classification and analysis~\cite{zhao2021plug-and-play}.  Meanwhile, the spatially related dependency is a point to enrich the inputted information and enhance the network comprehension of EEG data. EEG signals are strongly correlated temporal-spatial with long-dependency digital information. The signals belonging to different intentions are supposed to have rich and discriminative information. Therefore it is critical in BCI to explore the implicit relationship between EEG signal dimensions.

In the era of connected intelligence~\cite{LetaiefICM2019}, the proliferation of ubiquitous IoT devices has generated unprecedented volumes of multimodal data, which presents a new challenge and performance bottleneck for conventional communications~\cite{HouMobiCom2019}.

According to the three levels of pervasive communication proposed by Weaver and Shannon~\cite{shannonUIP1997}, it is necessary to address the semantic and security problem of communication under a data-centric scenario to make sure that the encrypted privacy-related data accurately expresses the expected meaning. Conventional communications encode the source into a series of meaningless bits ignoring the rich semantic information contained in the source. Different from technical problems, encoding-decoding communication has emerged recently to make use of the semantic information contained in the source, encoding the transmitted privacy-related data~\cite{strinatiCN2021}. By drawing the meanings from data and filtering out the redundant, irrelevant, and unimportant information, encoding-decoding communications can process data in the semantic domain, further compressing the data while reserving the meanings \cite{XieTSP2021}. There are various pieces of research on feeding text, voice, images, and other data as input to encoding-decoding communication frameworks. For example, DeepSC~\cite{KurkaTWC2021} was based on a transformer for text information transmission and distinguished semantic information at the sentence level for the first time. DeepSC-S~\cite{WengIJSAC2021} extended DeepSC to speech signal transmission and designed a semantic-based autoencoder based on the attention mechanism. To the best of our knowledge, The development of an encoding-decoding communication framework for EEG-based BCI has yet to be addressed. Consequently, numerous challenges and obstacles remain in modeling the system and require resolution.

In this paper, we aim to provide an accessible end-to-end system namely \textbf{BRIEDGE}, to demonstrate the feasibility of using civil EEG devices to control robots. It is a multi-brain to multi-robot interactive system through EEG-adaptive neural networks and encoding-decoding communications with a special loss function, as shown by the example in Fig.~\ref{fig_abstract}. BRIEDGE is specifically designed to control multiple intelligent agents simultaneously from multiple users wearing various low-cost EEG equipments and to solve the data obfuscation and security problems caused by heterogeneous data~\cite{xue2022imwut} generated from different types of EEG collection devices.

\subsection{Motivation Study}
\label{sec:motivation}
As the standardization of 5G-Advanced progress, a common theme in many perspectives is that 6G should be capable of integrating sensing, computing, communication and control~\cite{PanVTM2022,SuMobiCom2020}. It has been visioned that edge infrastructure can provide site-specific services for surrounding users, where the sensing, computing, communication, and control functionalities are simultaneously performed via a unified framework to optimize the performance since two main advantages can be acquired with this systematic operation, which is the integration gain realized through a synergistic design and coordination gain obtained via a co-design perspective. Considering existing BCI works usually focus on using EEG to control or interact with one specific robot; multi-brain to multi-robot interaction has been rarely tackled and still needs system design efforts to integrate sensing, computing, communication, and control. The crucial point in addressing the challenge of multi-brain to multi-robot lies in the development of a systematic algorithm. 
\begin{figure}[htb]
\begin{center}
\begin{tabular}{cc}
\includegraphics[width=3.5cm]{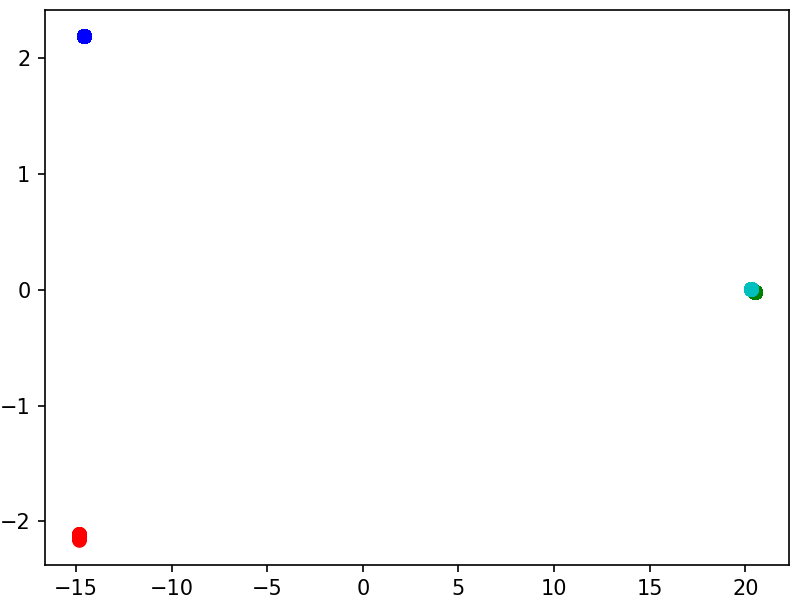} & \includegraphics[width=3.5cm]{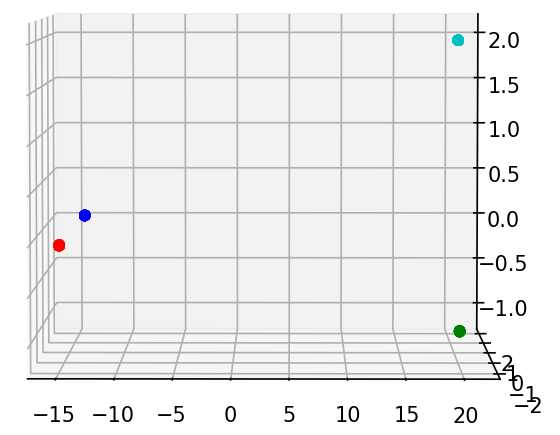} \\
{\scriptsize(a) Cluster result printed in 2D.} & {\scriptsize (b) Cluster result printed in 3D.} 
\end{tabular}
\begin{tabular}{c}
\includegraphics[width=8cm]{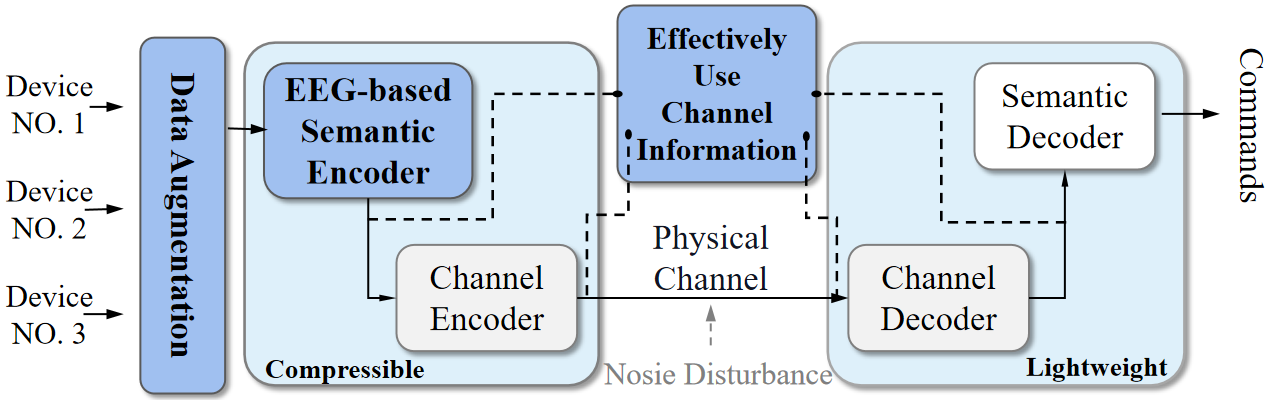} \\
{\scriptsize (c) \textcolor{black}{Exemplified framework of encoding-decoding communication framework.}}
\end{tabular}
\vspace{-4mm}
\caption{\textcolor{black}{Heterogeneous EEG data influence and encoding-decoding communication framework for EEG-based semantic information.}}
\label{fig:motivation}
\end{center}
\end{figure}
However, the lack of a standardized production system for brain-machine devices has resulted in a plethora of device variations available in the market, each exhibiting distinct levels of signal clarity and dimensionality when it comes to sensing.  The non-standardization of brain-computer devices presents the daunting challenge of effectively processing and recognizing heterogeneous data. In our motivation study, we utilized EEGNet~\cite{LawhernJNE2018} algorithm with a dimensional fusing module (composed of linear layers and convolutional layers) and attempted to extract features from four heterogeneous datasets (the dimensions of the data are [15, 4], [14, 1], [640, 64], and [80, 80], respectively), resulting in poor classification accuracy. \textcolor{black}{To identify the underlying cause of this problem, we visualized the features after the dimensional fusing module. The cluster groups of 4 heterogeneous data are shown in Figure~\ref{fig:motivation} (a) (b), revealing a noticeable gap between the lower- and higher-dimensional data. Here, the lower-dimensional data of [15, 4] and [14, 1] are marked as light green and green, whereas the higher-dimensional data of [640, 64] and [80, 80] are marked as red and blue, respectively. The noticeable gap suggests that the model can proficiently classify data from various dimensions. However, it cannot accurately classify specific EEG signals, which have been gathered in one dot, because the distinctions between EEG signals are unobservable. This inspired us to distinctively process the diverse dimension data to render the distance between EEG signals observable. As the 64 channels commonly be the boundary of medical and civil EEG devices, we defined higher- and lower-dimensional data based on whether the one data sample is greater than 4096 (64 $\times$ 64) or not in our paper.}

To ensure the security of the brain commands transmitted to the robots, the use of encoder-decoder architectures is a commonly employed approach~\cite{le2019COMMUNI, alawad2022GCC}. However, using civil EEG devices suffers from extra noise, challenging the accuracy. To address these issues, a multi-brain to multi-robot system is necessary to classify heterogeneous data with noise and extract precise semantic information from long-encoded commands. To improve the system's adaptive capacity in processing EEG data, we draw inspiration from previous studies on data augmentation~\cite{zhang2018icml, nagata2022SP} and transmission enhancement techniques~\cite{dong2022semantic, uysal2022semantic}. Taking these issues into consideration, we integrated the mechanism of encoding-decoding communication as exemplified in Figure~\ref{fig:motivation} (c), and proposed a paradigm shift to a semantic-level communication network, which offers comparable quality of service and data security while transmitting data to Edge AI~\cite{niu2022paradigm}. However, encoding-decoding communication faces an accuracy problem during the transmission compared with traditional communication, we try to design a suitable loss function to improve the transmission accuracy of our framework.

\subsection{Challenges}
There are three major challenges to realize BRIEDGE. \textit{\large{The first challenge}} comes with its inherent nature of multi-brain to multi-robot interaction scenario, where various collection devices on multiple users present heterogeneous even multimodal data.
 
In addition, there may be a coexistence of non-invasive BCI and invasive BCI, which can concurrently present multimodal signal features at the same time. Therefore, an EEG-adaptive neural network structure is desirable for semantic feature extraction and classification. However, traditional neural networks for EEG data have their disadvantages. For example, RNNs suffer from limited parallel computing capabilities and time-consuming computations due to the sequential structure. CNN with the capability of parallel computing can use deeper network to extract EEG-based semantic information from long sentences, but its performance is not as good as that of RNN because the kernel size in CNN is small to guarantee computational efficiency.

\textit{\large{The second challenge}} follows the first one where multi-brain to multi-robot interaction resulting in heterogeneous sources and enormous amounts of data raises reliability and data security issues in end-to-end communication. Because signal transmission between numerous EEG devices and robots requires a variety of inputs, multiple channels and diverse noises, the channels and noises are continually changing over time. Although the emergence of encoding-decoding communication could be a promising way to solve this problem, multimodal data transmission from multiple devices is still rarely tackled. To address the parallel transmissions under multi-user and multi-task scenarios, an encoding-decoding communication framework needs to define semantic information of EEG signals and extract semantic features of EEG signals at the transmitter for different EEG devices, finally encrypting them. In addition, given the low fidelity and noisy transmission, BRIEDGE is expected to combine the 
 EEG-based semantic information extraction method with the end-to-end channel to realize the end-to-end autoencoder and autodecoder modeling of encoding-decoding communication. To improve the encoder-decoder communication, a loss function based on the actual system is particularly important. Throughout this process, data security is well protected due to the deployment of an encoder-decoder structure.


\textit{\large{The third challenge}} is relevant to the mobile computing and networking requirement for multi-brain to multi-robot interaction. Edge AI~\cite{FangMobiCom2018,HanMobiCom2021,ZhangIMWUT2020} could play a key role to meet the agile needs in BRIEDGE. Simply put, edge AI is a combination of edge computing and artificial intelligence, where AI algorithms are processed locally (either directly on the device or on the server near the device) so that devices can make independent decisions. The transmitter running autoencoder model in BRIEDGE could be either an edge mobile server or an edge portable server similar to the Apple TV box~\cite{AppleTV} but usually with resource constraints on hardware. Also, the receiver running the autodecoder model is usually deployed in mobile robots with insufficient computing capability. Therefore, in BRIEDGE, the incorporation of specific model compression techniques~\cite{HanICLR2016} is expected. This process can accelerate neural network inference while minimizing the consumption of computing resources without affecting the loss of semantic information.

\subsection{Contributions}
To the best of our knowledge, BRIEDGE is the first system that enables multi-brain to multi-robot interaction through EEG-adaptive neural networks and encoding-decoding communications framework. Specifically, to tackle the first challenge, we introduce an intermittent masked mechanism and an Informer-based ProbSparse self-attention mechanism to extract the joint features of heterogeneous EEG data and then use dynamic feature integration to improve the classification accuracy. To address the second challenge, we design a dynamic EEG-based semantic autoencoder enhanced with a brain transformer and corresponding semantic autodecoder to inherit the EEG-based semantic information from the transmitter. Also, an EEG-based semantic performance metric is jointly modeled by the mutual information function to measure the channel codec performance and the cross entropy function to measure the EEG-based semantic codec performance. In terms of the third challenge, model compression schemes with pruning, weight sharing, and quantization are deployed to support the transmitter running an EEG-based semantic encoder model on a resource-constrained edge portable server. A lightweight channel decoder and EEG-based semantic decoder with residual technique and fully connected layer are also deployed on the receiver side of the robot. In addition, we provide a code map representing various commands for multiple users to control multiple intelligent agents. In our paper, BRIEDGE demonstrates the feasibility of using civil EEG devices to control robots and reveals the potential application of controlling robots via low-cost EEG devices.

\subsection{Applications and Impacts}
Metaverse is emerging as the most promising platform for future Internet, where BCI represents one of the key enabling technologies in metaverse~\cite{LeeARXIV2021}. In comparison with invasive BCI that is implanted directly into the brain during a neurosurgery, non-invasive BCI mostly working on the principles of EEG with comparably lower price has the potential for a better market penetration rate. Currently, our BRIEDGE is focused on non-invasive BCI and multi-brain to multi-robot interaction, which were barely tackled in previous works but good for metaverse scenarios such as co-creation~\cite{LouieCHI2020}, autonomous manufacturing or unmanned factory~\cite{XiongTCSS2022}, and Immersive technology or extended reality (XR)~\cite{LiuMobiCom2019}, with massive access and control under an increasing number of intelligent agents and human-machine-thing interaction needs. Therefore, the scenario defined in BRIEDGE has meaningful implications for both research and practice. Meanwhile, the EEG-adaptive neural networks as an EEG-based semantic encoder and encoding-decoding communications in BRIEDGE present a generic system for brain signal analysis and feedback on heterogeneous or multimodal data generated from various sensing devices. Its design and optimization schemes give a reference case integrating sensing, computing, communication, and control, which could benefit industry and academy communities working towards 6G and beyond~\cite{PanVTM2022,SuMobiCom2020}.

\section{System Overview}

\begin{figure*}[h]
	\centering
	\includegraphics[width=0.9\linewidth]{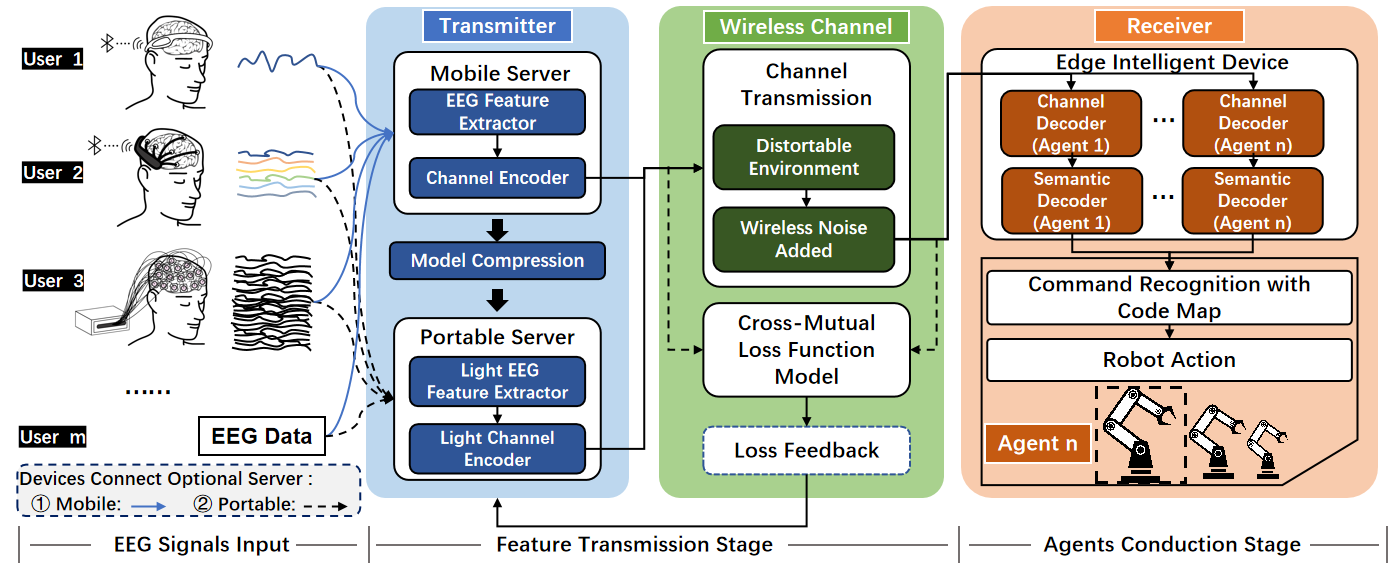}
	\vspace{-5mm}
	\caption{The end-to-end system architecture of BRIEDGE.}
	\label{fig1}
\end{figure*}

The system operation of BRIEDGE is simply described as follow: a variety of EEG devices gather EEG signals, transmit them to an edge mobile server or an edge portable server for encoding, and then send the encoded signals to several robots for decoding and executing the instructions. Fig.~\ref{fig1} illustrates its end-to-end system architecture composed of EEG signal collection stage, feature transmission stage and agent execution stage. 

In the EEG signal collection stage, various users wear EEG wearable headsets with different number of electrodes to capture their EEG signals and transmit to the transmitter module. Each raw EEG data is collected within a customized sampling interval and represented in decimal format. Since the devices collecting EEG data are different, the channel number and feature dimension of each type of EEG data vary as well. These collected datasets are labeled and divided into different categories based on the types of brain activity. Then the EEG data is fed into a pre-processing algorithm for down-sampling and augmentation with Gaussian noise.

The feature transmission stage can be split into a transmitter module and a wireless channel module. Transmitter is responsible for feature extraction and signal transmission, which can be deployed in an edge mobile server with high computational resources or an edge portable server (embedded unit) with less computational resources. EEG signals recorded from different EEG devices are sent the server first, and then BRIEDGE employs a dynamic encoder to obtain the EEG-based semantic feature for generating channel encoder. Model compression is also used to achieve lightweight feature encoder and channel encoder deployed in edge portable server. Through multiple iterations, the dynamic EEG feature encoder can adapt to different types of EEG data to achieve better signal recognition. In the wireless channel module, BRIEDGE transmits the encoded EEG semantic features into different channel environments with respective Gaussian noise. A proposed cross-mutual loss function is used to calculate the correlation between the encoded and decoded features. After several iterations, the EEG-based semantic information of encoded features and decoded features will become consistent to avoid information loss during transmission.

The agent execution stage is performed when the EEG signal transmitted through the wireless channel arrives at the receiver module, which is embedded in each robot and includes edge intelligent agents, a command code map, and robot action. The signals at the receiver are decoded by semantic decoders and then the decoded signals are decompressed by channel decoders. Next, BRIEDGE maps the recognized signals to the corresponding command sent to the designated robot based on a code map. After that, the robot performs a specified action to complete the EEG task following the command.

\section{Design of BRIEDGE}


\textcolor{black}{BRIEDGE consists of an EEG-adaptive neural network named dynamic EEG feature encoder in the transmitter, and an encoding-decoding communication framework across the transmitter, channel transmission, and receiver.} We illustrate the design and implementation of essential modules and functions of BRIEDGE as depicted in Fig.~\ref{fig1}, to address the challenges. 

\subsection{Dynamic EEG Feature Extractor}\label{3.1}
EEG data collected from different devices has issue of heterogeneous data forms that must be addressed. We devise a dynamic encoder that adaptively extracts heterogeneous features, aiming to improve the understanding of the heterogeneous EEG data to achieve an accurate performance of BRIEDGE. \textit{Note that the dynamic EEG feature encoder here also served as the semantic encoder in Section~\ref{semantic_section}.}

\subsubsection{\textcolor{black}{Dynamic Path Selection}}
BRIEDGE processes EEG data separately based on their corresponding dimensions from different devices. In Fig.~\ref{fig:DE}, low-dimensional data (green and red boxes) and high-dimensional data (blue boxes) are differentiated. During training or testing, our dynamic feature encoder intelligently selects feature learning paths based on input data dimensionality. Low- and high-dimensional data are embedded into [128, 128] dimensions with added Gaussian noise. Then, intermittent mask augmentation and joint feature extraction are applied. The encoder retains the gradient of the appropriate path and freezes the other based on data dimensionality. This dynamic process allows the dynamic adjustment of network weights for various EEG devices, with minimal path selecting time. Once the data from a device has been trained, it can be switched to use that device at any time in the system without retraining it.

\subsubsection{\textcolor{black}{Dimensional Feature Encoder}}

\begin{figure*}[tb]
    \begin{minipage}{0.5\textwidth}
        \centering
        \includegraphics[width=\linewidth]{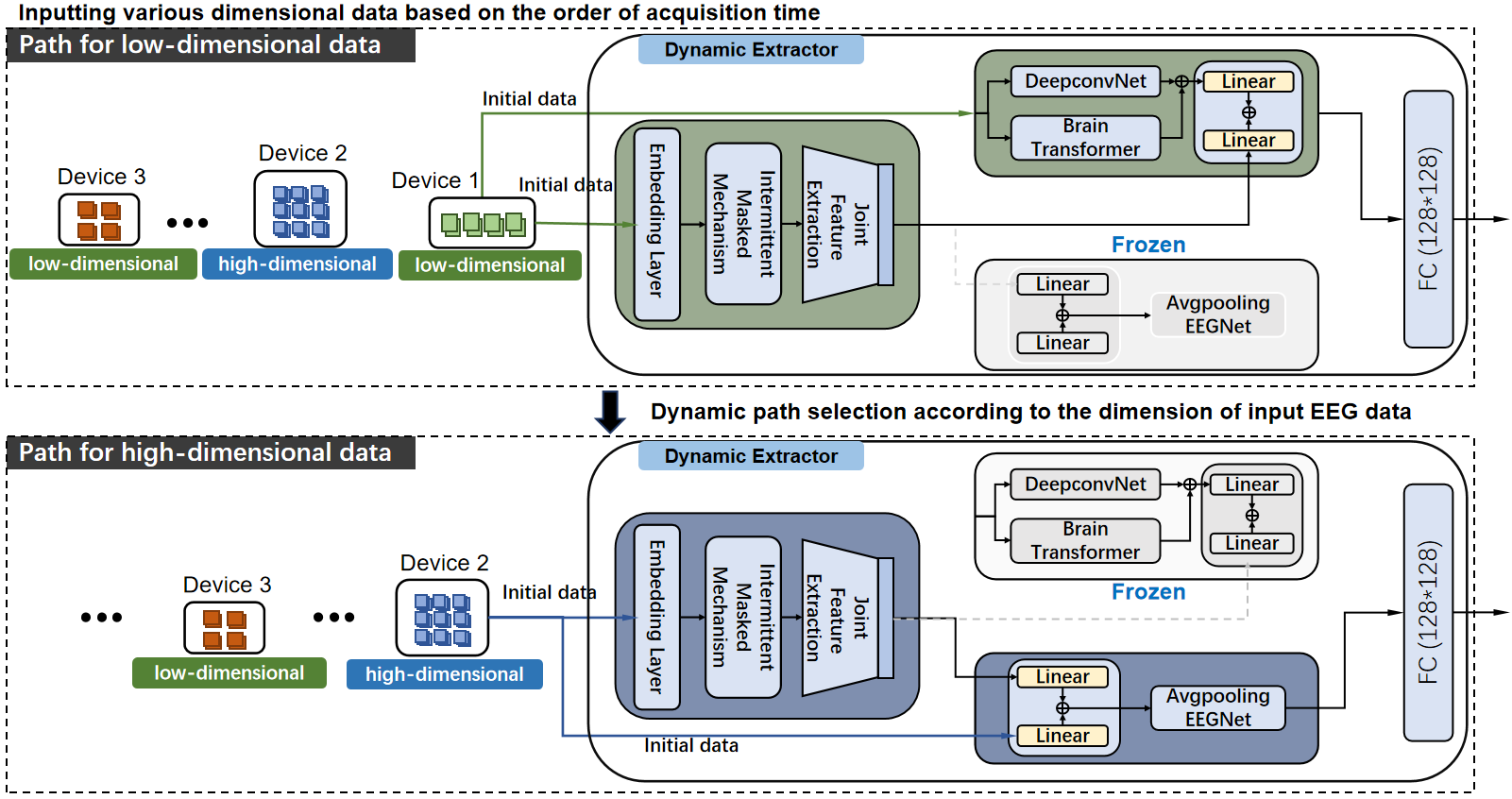}
        \vspace{-7mm}
        \caption{\textcolor{black}{Dynamic EEG feature encoder network.}}
        \label{fig:DE}
    \end{minipage}%
\hfill
    \begin{minipage}{0.5\textwidth}
        \centering
        \includegraphics[width=\linewidth]{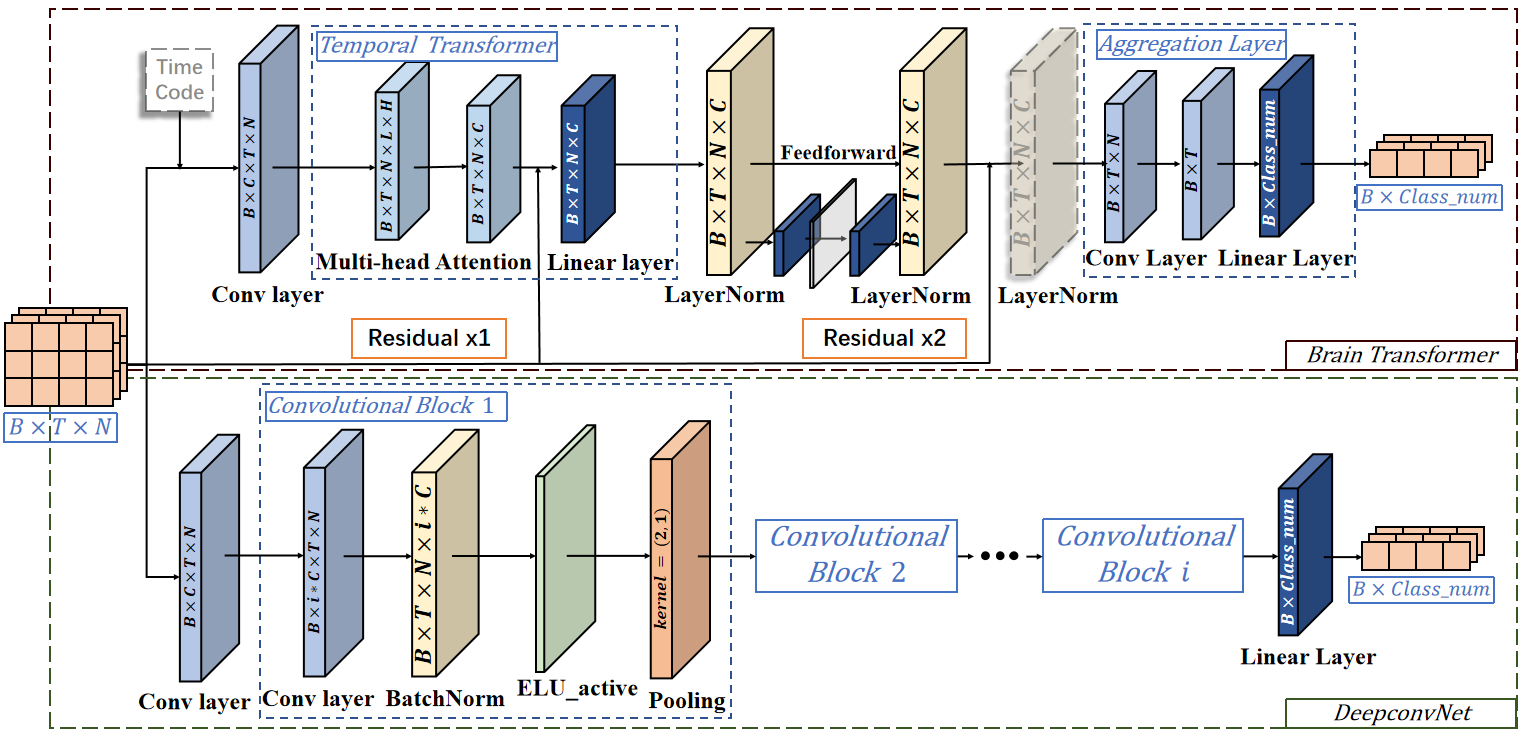}
        \caption{\textcolor{black}{Integrated module for low-dimensional data.}}
        \label{fig_BT}
    \end{minipage}
    \vspace{-4mm}
\end{figure*}

\textcolor{black}{As shown in Fig.\ref{fig:DE}, low- and high-dimensional data follow different feature extraction paths. We integrated EEGNet~\cite{LawhernJNE2018} for high- dimensional EEG data, brain transformer~\cite{WuUbiComp2022} and DeepconvNet~\cite{SchirrmeisterHBM2017} for low-dimensional high-noise data. Each model was accordingly optimized, where details are provided in our code. For brain transformer, we increased the heads of attention mechanism to 8 and reconstructed the convolutional layers with residual structures. Then DeepconvNet gained more convolutional blocks and added residual structures, while EEGNet's pooling layer was updated for better feature integration. From Fig.\ref{fig_BT}, we illustrate the incorporation of the brain transformer and DeepconvNet. This structure enhances the capacity to capture temporal dependencies in low-dimensional EEG data after embedding. }

To prevent the potential loss of EEG features during learning, we introduced a weight assignment model for proportional fusion of initial data and features (see Fig.\ref{fig:DE}). The assignment model, which involves 2 linear layers, automatically assigns weights for features. Two assignment models are applied for the front and back fusion of high- and low-dimensional encoders. Before the data is fed into a particular path, we introduce a feature extraction approach, comprising an intermittent mask mechanism and a joint feature extraction module, to further enhance feature extraction. \textcolor{black}{More detailed improvement can be observed in our code.}

\begin{figure}[htb]
  \begin{minipage}[c]{1\linewidth}
    \centering
    \includegraphics[width=0.95\columnwidth]{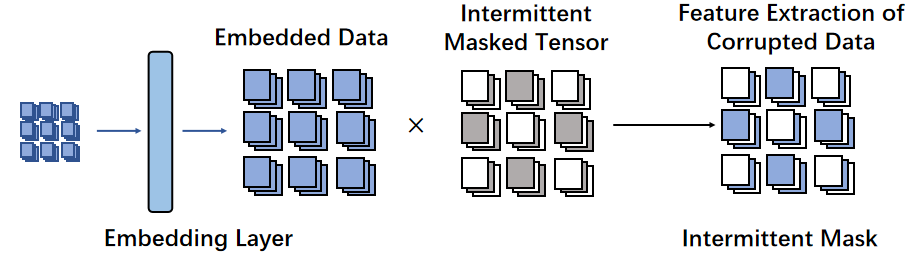}
	\vspace{-2mm}
	\caption{Intermittent Masked Mechanism.}
	\label{fig_mask}
  \end{minipage}%
  \\
  \begin{minipage}[c]{1\linewidth}
    \centering
    \includegraphics[width=0.95\columnwidth]{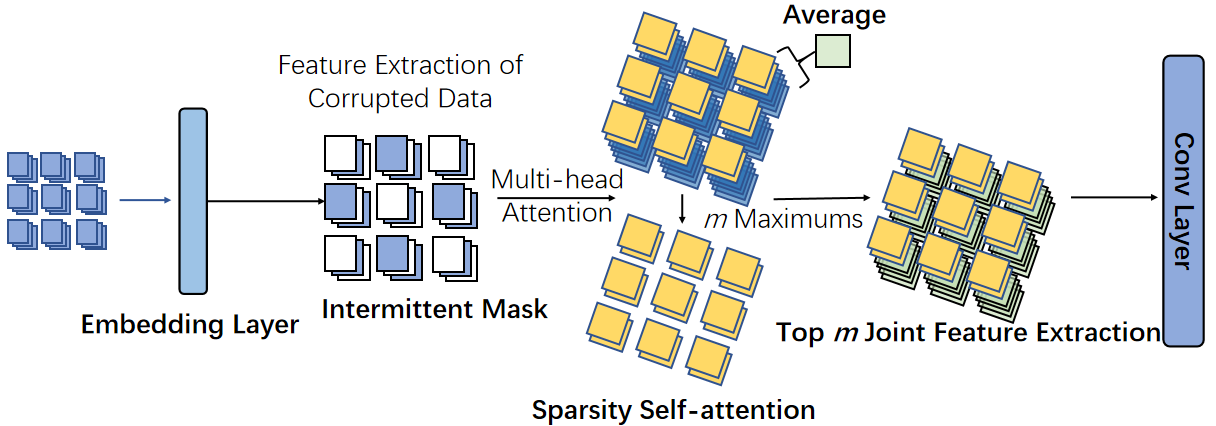}
	\vspace{-2mm}
	\caption{Joint Feature Extraction.}
	\label{fig:JFE}
  \end{minipage}%
\end{figure}

\textbf{Intermittent Mask Mechanism.}
\textcolor{black}{Brain signals collected at a single moment are less reliable due to noise in the collecting process. Collecting EEG data over time provides a more accurate representation of brain activity. However, the EEG data have some trend-related representations in the temporal dimension that need attention. The intermittent mask tensor of adjacent regions could enhance the abilities of feature learning~\cite{liicip2022}. Therefore, we proposed an intermittent mask mechanism presented in Fig.~\ref{fig_mask} for EEG data encoding to learn the feature representation of trend-related dependencies. This mechanism employs a zero and one mask tensor to modify input data, highlighting data trends and mitigating the disturbances between adjacent data points in the temporal dimension. The intermittent mask mechanism simulates data interference caused by noise or sensor disconnect in non-invasive EEG devices, addressing classification accuracy issues associated with the first challenge.}

\textbf{Joint Feature Extraction Module.}
\textcolor{black}{EEG data from diverse devices, despite exhibiting heterogeneous formats, possess joint features that can be effectively addressed through the utilization of the self-attention mechanism. To more efficiently extract the joint features, we adopted some cutting-edge methods to reinterpret the attention mechanism for efficient extraction. Previous studies~\cite{Child2019arxiv,Beltagy2020arxiv,Li2019nips,zhouAAAI2021} have discovered that the distribution of self-attention probabilities is potentially sparse and inspired the improvement of the sparsity self-attention mechanism, which can capture the long-range dependencies within long-sequence time-series inputs while reducing time complexity and memory usage. 
Therefore, we consolidate these sparsity self-attention mechanisms into this module for the efficient extraction of joint features.}

\textcolor{black}{The traditional self-attention mechanism requires the quadratic time's dot-product computation and $\mathcal{O}(L_{Q}L_{K})$ memory usage, which increases the computational difficulty of capturing complicated joint features in edge devices. According to the previous works, the contribution of each dot-product result in the attention mechanism follows two distributions in the self-attention mechanism query matrix: a normal distribution with a more average contribution and a dominant distribution with a more active contribution. We select Top-$m$ active queries to represent the top contributing joint features in our module. Then, the Kullback-Leibler divergence is introduced to describe the likeness between distributions for further enhancement. $m$ could be set as nearly 25\% of the $L_{Q}$ and we set $m=30$ in our system. The detailed structure can also be observed in our code.}

Then we greatly reduce memory usage to $\mathcal{O}(\ln{L_{Q}} L_{K})$ of our sparsity self-attention. Following behind the intermittent mask mechanism, the sparsity self-attention mechanism would select and remain the top $m$ dominant dot-product pairs then recover the rest of the pairs with average. The joint features from different devices' intermittent masked data would subsequently be learned by a convolutional layer, which could enhance the model's understanding of trend-related characteristics across heterogeneous EEG data.
\subsubsection{Model Compression}~\label{compression} \textcolor{black}{Model compression accelerates neural network inference and reduces power consumption by employing techniques such as pruning, quantization, and knowledge distillation~\cite{HanICLR2016}, which benefit the edge computation. In BRIEDGE, pruning and quantization are utilized for model compression. We first use weight pruning (unstructured pruning), which involves setting a threshold to remove low-magnitude weight values. It computes the standard deviation for weight matrices after training and prune values below a threshold (0.25 times the standard deviation) to 0, creating sparse matrices stored in CSR or CSC format. Subsequently, filter pruning (structured pruning) is utilized to remove filters with negligible effect on loss using Taylor expansion~\cite{YouNeurIPS2019}, maintaining performance with fewer parameters and mitigating overfitting~\cite{LeCunNIPS1989}.}

\textcolor{black}{After that, we use two quantization methods: weight sharing and half-quantization. The weight-sharing method clusters weight values with $k$-means, using cluster indices to represent similar values. The compression rate $r$ is calculated as $\mathop r=\frac{n b}{n \log _{2}(k)+k b}$, with $n$ connections, $b$ bits per weight, and $k$ clusters. Then, half-quantization converts the stored model from Float32 to Float16 after weight sharing. This combined approach optimizes the model size and accelerates the inference process.}

\subsection{Channel Transmission}

\subsubsection{Encoding-Decoding Communication Process}\label{semantic_section}
The end-to-end transmission mostly operates in a downlink for BRIEDGE system, where line-of-sight (LOS) communication is ensured for both the user-server and server-robot links. Assuming the $m$ user-server links are available, we only model the server-to-robot transmission as an encoding-decoding communication module.

\begin{figure}[htb]
	\centering
	\includegraphics[width=1\columnwidth]{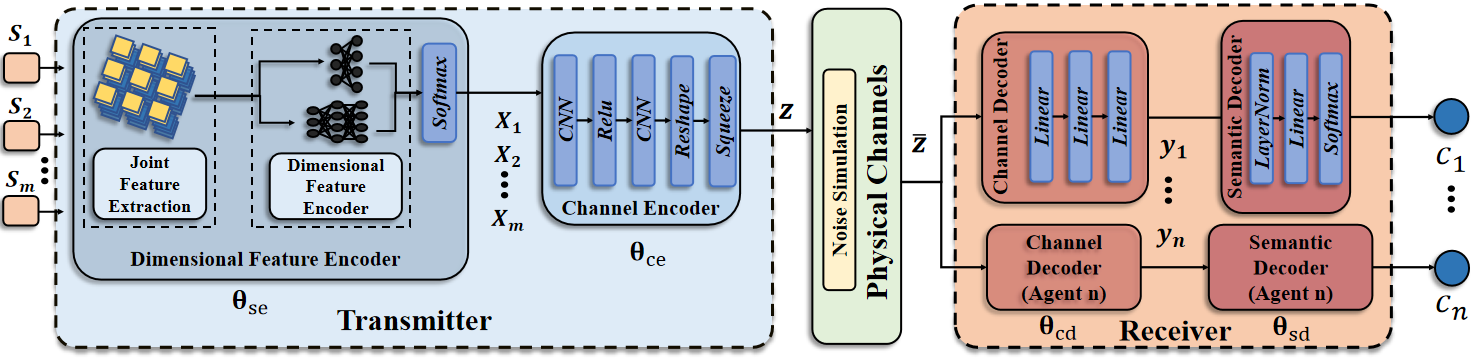}
	\vspace{-2mm}
	\caption{The structure of the encoding-decoding communications framework for precisely extracting and transmitting the EEG-based semantic information and interpreting it into commands}
	\label{fig2}
\end{figure}
\textbf{Transmitter.} The encoding-decoding communication module in BRIEDGE is modeled in Fig.\ref{fig2}. The input of the transmitter is an EEG signal ${s}_{i}$, representing the i-th EEG device's signal, and $i \in [1,2,...,m]$ where $m$ represents the connected users. The transmitter consists of a dimensional feature encoder (served as an EEG-based semantic encoder) and a channel encoder, which are used to extract the EEG-based semantic features from ${s}_{i}$ and compress them for the physical channel transmission. The input ${s}_{i}$ is mapped to ${x}_{i}$ by the semantic encoder and then transmitted by channel encoder. ${x}_{i}=f_{\boldsymbol{\theta}_{s e}}({s}_{i})$ is a function of semantic encoder, where $\boldsymbol{\theta}_{s e}$ is represented as the parameters of semantic encoder networks. In addition, the specific structure of the semantic encoder can refer to Section \ref{3.1}. Then ${x}_{mm}=[{x}_{1}, {x}_{2},..., {x}_{m}$] is viewed as the input of channel encoder. Similarly, $z=f_{\boldsymbol{\theta}_{c e}}({x}_{mm})$ is a function of channel encoder, where $\boldsymbol{\theta}_{c e}$ is represented as the parameters of channel encoder networks.  Here, the network parameters of the transmitter is denoted as $\boldsymbol{\theta}^{\mathcal{T}}=(\boldsymbol{\theta}_{s e}, \boldsymbol{\theta}_{c e})$.

\textbf{Channel.} 
\textcolor{black}{The transmission process from the transmitter to the receiver can be modeled as:
$\bar{z}=h * z+w$, 
where $w \sim \mathcal{C N}\left(0, \sigma^{2} \mathbf{I}\right)$ is independent and identically distributed Gaussian noise, $\sigma^{2}$ is the noise variance for each channel, $h$
is the coefficients of a linear channe, and $\mathbf{I}$ is the identity matrix.}

\textbf{Receiver.} The channel decoder and the semantic decoder are cascaded to act as a receiver. They are not complex in structure and require only a small amount of computing resources for deploying. The construction of receiver could be observed in Table.~\ref{tab:receiver}. The semantic decoder recovers EEG commands based on the learned and retrieved EEG-based semantic features, while the channel decoder reduces channel distortion and attenuation through parameters convey in latent space. Similarly, we denote the network parameters of the channel decoder and the semantic decoder as $\boldsymbol{\theta}_{c d}$ and $\boldsymbol{\theta}_{s d}$ respectively. As depicted in Fig.~\ref{fig2}, the decoding command of robot arm $c_n$ can be obtained from the received signal $\bar{z}$, by the following operation: $c_n=f_{\boldsymbol{\theta}_{s d}}(f_{\boldsymbol{\theta}_{c d}}(\bar{z}))$, where $f_{\boldsymbol{\theta}_{c d}}$ and $f_{\boldsymbol{\theta}_{s d}}$ indicate the channel decoder and semantic decoder respectively. Here we denote the network parameters of the receiver as $\boldsymbol{\theta}^{\mathcal{R}}=(\boldsymbol{\theta}_{c e}, \boldsymbol{\theta}_{s e})$.

\subsubsection{Cross-Entropy and Mutual Loss Function}
Considering that the performance of BRIEDGE is mainly affected by the semantic encoding and decoding module, channel encoding and decoding module, and channel transmission module, we proposed a cross-entropy and mutual information loss function to enhance the accuracy.

\textbf{The First Training Phase.}
Data transmission rate and channel capacity are important performance indexes of communication framework. We can measure the correlation of the two indicators by mutual information. The smaller the mutual information is, the correlation of the two indicators is smaller.
The mutual information of the channels' input $z$ and the channels' output $\hat{z}$ can be represented by $I(z ; \hat{z})$.
The mutual information can also be expressed as the KL divergence of the product $p(\cdot) \times p(\cdot)$ of the marginal distributions of two random variables $z$ and $\hat{z}$ relative to the joint entropy $p(z, \hat{z})$ of the random variable, which is represented as $\label{eq:Shi3} I(z ; \hat{z})=D_{\mathrm{KL}}(p(z, \hat{z}) \| p(z) p(\hat{z}))$.

\textcolor{black}{We then find a tight lower bound on the $I(z; \hat{z})$, and an unsupervised method is used to train the network $T$, where the expectation can be calculated by sampling from the above KL divergence~\cite{belghaziICML2018,handschin1969monte}. After that, we can optimize the encoder by maximizing the mutual information defined in the above KL divergence. The related loss function can be given by:
\vspace{-2mm}
\begin{equation}
\small
\mathcal{L}_{\mathrm{MI}}(z, \hat{z} ; T)=\mathbb{E}_{p(z, \hat{z})}\left[f_{T}\right]-\log \left(\mathbb{E}_{p(z) p(\hat{z})}\left[e^{f_{T}}\right]\right)
\end{equation}
where $f_{T}$ is composed by a neural network, in which the inputs are samples from $p(z, \hat{z})$, $p(z)$ and $p(\hat{z})$.} \textcolor{black}{In our function, $z$ is generated by the function $f_{\boldsymbol{\theta}_{s e}}(\cdot)$ and $f_{\boldsymbol{\theta}_{c e}}(\cdot)$, and the loss function can be used to train neural networks to get $\boldsymbol{\theta}^{\mathcal{T}}$ and $T$.}

\textbf{The Second Training Phase.}
In the second phase, the cross-entropy is used as the loss function to measure the difference between the channel encoder's input ${x}_{m}$ and the channel decoder's output ${y}_{n}$ as follows:
\vspace{-2mm}
\begin{equation}
\scriptsize
\mathcal{L}_{\mathrm{CED}}({x}_{m}, {y}_{n} ; \boldsymbol{\theta}^{\mathcal{T}}, \boldsymbol{\theta}_{c d})= -\sum_{l=1}^{L}[ q\left(o_{l}\right) \log \left(p\left(o_{l}\right)\right)+\left(1-q\left(o_{l}\right)\right)\log \left(1-p\left(o_{l}\right)\right)]
\label{eq:CED_loss}
\end{equation}
where $q\left(o_{l}\right)$ is the discrete probability distribution that the $l$-th term $o_{l}$, appears in the estimated signal $x_m$, and $p\left(o_{l}\right)$ is the predicted probability distribution that the $l$-th term $o_{l}$, appears in the signal $y_n$.

\textbf{The Third Training Phase.}
In the final phase, the cross-entropy is used as the loss function to measure the difference between the semantic encoder's input ${s}_{m}$ and the semantic decoder's output ${c}_{n}$ as follows:

\begin{equation}
\scriptsize
\begin{aligned}
\mathcal{L}_{\mathrm{SED}}({s}_{m}, {c}_{n} ; \boldsymbol{\theta}^{\mathcal{T}}, \boldsymbol{\theta}^{\mathcal{R}})=-\sum_{l=1}^{L}[ q\left(o_{l}\right) \log \left(p\left(o_{l}\right)\right)+\left(1-q\left(o_{l}\right)\right)\log \left(1-p\left(o_{l}\right)\right)]
\end{aligned}
\label{eq:SED_loss}
\end{equation}
where $q\left(o_{l}\right)$ is the discrete probability distribution that the $l$-th term $o_{l}$, appears in estimated signal $s_m$, and $p\left(o_{l}\right)$ is the predicted probability distribution that the $l$-th term $o_{l}$, appears in sentence $c_n$.

Thus, the loss function of BRIEDGE can be expressed as 
\begin{equation}
\scriptsize
\begin{aligned}
\mathcal{L}_{\mathrm{C-M}}(\boldsymbol{\theta}^{\mathcal{T}},\boldsymbol{\theta}^{\mathcal{R}}) &=\mathcal{L}_{\mathrm{SED}}({s}_{m}, {c}_{n} ; \boldsymbol{\theta}^{\mathcal{T}}, \boldsymbol{\theta}^{\mathcal{R}}) +\lambda\mathcal{L}_{\mathrm{CED}}({x}_{m}, {y}_{n} ; \boldsymbol{\theta}^{\mathcal{T}}, \boldsymbol{\theta}_{c d})
\\-\gamma\mathcal{L}_{\mathrm{MI}}(z, \hat{z} ; T, \boldsymbol{\theta}^{\mathcal{T}})
\end{aligned}
\label{eq:mix_loss}
\end{equation}
\textcolor{black}{where $\lambda$ and $\gamma$ $(\lambda,\gamma \in (0,1])$ are the weights for the second term and the third term, respectively. By using this three-stage loss function, BRIEDGE achieved 24\% accuracy enhancement compared with just using the third training phase of the loss function. }

\subsection{Interaction and Execution}

\subsubsection{Multi-Brain to Multi-Robot Interaction} 
\begin{figure*}[htb]
\begin{center}
\includegraphics[width=1.8\columnwidth]{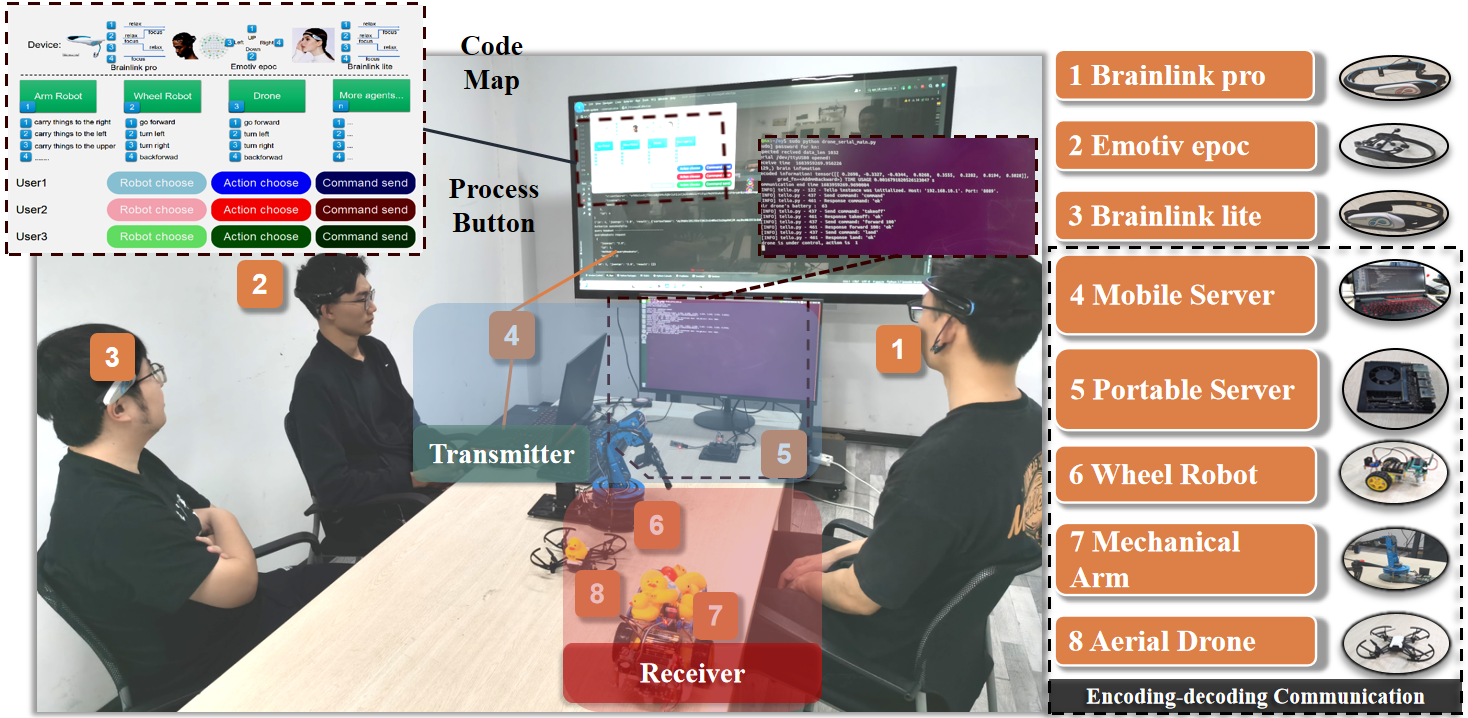}
\caption {\textcolor{black}{BRIEDGE deployment and testing scenario for multi-brain to multi-robot interaction.}}
\label{fig:interaction_system}
\end{center}
\end{figure*}

The multi-brain to multi-robot interaction scenario is illustrated in Fig.~\ref{fig:interaction_system}. In this testing case, three users wearing different types of EEG collection devices to control and coordinate three types of robots to complete various tasks simultaneously through end-to-end parallel commands. For example, user wearing Emotiv Epoc controls a wheel robot to transport an object to a nearby place, user wearing Brainlink Pro moves a robot's mechanical arm to grasp the object, and a user wearing Brainlink Lite put it on a drone robot to deliver the object to a remote place through EEG-based encoding-decoding communications.
Specifically, after EEG signals are successfully collected, the EEG signals are directly transmitted to a connected edge mobile server or edge portable server (embedded unit) as sequential and concurrent data via wireless Bluetooth serial port. After the EEG features are extracted and compressed at edge side, the command represented by the EEG signals will be displayed on a visual screen for users to judge whether the command can be transmitted to the robots. Users can delete this command if they do not wish to send it or if the edge side gets the incorrect command. When users receive a satisfactory command signal, they have the option of transmitting it from the edge side to the robot through Bluetooth. Based on the encoding-decoding communication framework explained in Section~\ref{semantic_section}, the encoded data with underlying EEG-based semantic information are transmitted to all edge robots at the same time. In this case, one or more commands could be executed by robots in parallel. When various EEG devices connect with the edge side and transmit over a wireless channel, robots hosting intelligent agents as the receivers are under different communication environments as well. This complex multi-path communication environment poses great challenges to the accurate transmission of EEG signals. When the command signal is transmitted to a robot, the robot needs to use its onboard resource to decode the command and execute the specific task according to a preset code map as shown in Section.~\ref{sssection:code_map}. The implementation of this process requires a lightweight design on both command coding and decoding mechanisms.

\subsubsection{Lightweight Decoder and Code Map}\label{sssection:code_map}

Considering the low computing capacity in robots, we designed lightweight models for the channel decoder and the semantic decoder deployed at the receiver side, wherein the channel decoder consists of three fully-connected layers, and two ReLU activation functions, and the semantic decoder is composed of a LayerNorm layer and a fully-connected layer as depicted in Table.~\ref{tab:receiver}. 

A code map defines the semantic information of the EEG signal in order to realize the multiple-agent strategy. The specific code map is set up for multiple agents to recognize the actual command from different users as shown in Table~\ref{tab:code map}. In a particular engagement for users, users should anticipate their expectations about which agents conduct what actions. Each agent has its own action list, such as a mechanical arm having a preset action list: [turn left, turn right, catch, put down] that can be selected to conduct. As for the engagement of new devices, corresponding new command categories will be stacked into the code map.  For example, user 1 with device 1 should first decide which agent to choose, then decide what action the agent should conduct that is listed in its action list. A visualized illustration for realistic interaction is shown in Figure~\ref{fig:interaction_system}. Subsequently, user 1 sends brain signals to first map a command to designate an agent ID and then map another command to choose an action ID. Finally, the agents' actors find their missions according to the command of the first position and then force agents to conduct actions according to the command of the second position. Therefore, this code map mechanism is good for controlling multiple agents from multiple EEG devices. 

\vspace{2mm}
\begin{minipage}[c]{0.45\textwidth}
\centering
\captionof{table}{BRIEDGE's Receiver Settings}
\vspace{-3mm}
\centering
\label{tab:receiver}
\includegraphics[width=0.9\columnwidth]{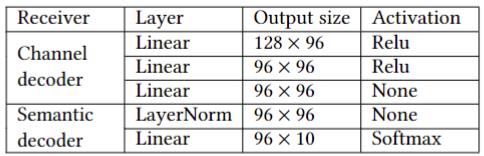}
\end{minipage}
\\
\begin{minipage}[c]{0.5\textwidth}
\centering
\small
\captionof{table}{Code Map Example}
\centering
\vspace{-3mm}
\label{tab:code map}
\setlength{\arrayrulewidth}{1pt}
\scalebox{0.75}[0.65]{$
\begin{tabular}{|c|c|c|c|c|}
    \hline
    \begin{tabular}[c]{@{}c@{}}Device\\ ID\end{tabular} & \begin{tabular}[c]{@{}c@{}}Category \\ Description\end{tabular} & \begin{tabular}[c]{@{}c@{}}Category \\ Code\end{tabular} & \textbf{\begin{tabular}[c]{@{}c@{}}First Position\\ (Agent ID)\end{tabular}} & \textbf{\begin{tabular}[c]{@{}c@{}}Second Position\\ (Action ID)\end{tabular}} \\ \hline 
    Device 1 & Totally foucs & Code 1 & ID 1 & Action 1 \\ \cline{2-5} 
     & Foucs to relax & Code 2 & ID 2 & Action 2 \\ \cline{2-5} 
     & Relax to focus & Code 3 & ID 3 & Action 3 \\ \cline{2-5} 
     & Totally relax & Code 4 & ID 4 & Action 4 \\ \hline
    Device 2 & Eyes opened & Code 5 & ID 1 & Action 1 \\ \cline{2-5} 
     & Eyes closed & Code 6 & ID 2 & Action 2 \\ \hline
    Device 3 & \begin{tabular}[c]{@{}c@{}}Fists \\ Open or close\end{tabular} & Code 7 & ID 1 & Action 1 \\ \cline{2-5} 
     & \begin{tabular}[c]{@{}c@{}}Imaging fists \\ open or close\end{tabular} & Code 8 & ID 2 & Action 2 \\ \cline{2-5} 
     & \begin{tabular}[c]{@{}c@{}}Feets \\ open or close\end{tabular} & Code 9 & ID 3 & Action 3 \\ \cline{2-5} 
     & \begin{tabular}[c]{@{}c@{}}Imaging feets \\ open or close\end{tabular} & Code 10 & ID 4 & Action 4 \\ \hline
    \end{tabular}$}
\end{minipage}

\section{Accuracy Evaluation on Heterogeneous Data}
\label{sec:expriment1}
\subsection{Datasets}
To preliminarily validate the generalization capability of our system on different EEG devices, we evaluate the accuracy of our system using four heterogeneous datasets: one self-collected dataset (Brainlink), two public datasets (EyeState~\cite{Dua2017}, BCI-2000~\cite{SchalkTBE2004}), and a mixed dataset of the above three (Hybrid).

\begin{itemize}
\item {\textbf{Brainlink~\cite{brainlink}}.} 
This dataset contains 800 EEG samples collected from 8 subjects using Brainlink Lite, with each subject's measurement duration lasting 180 seconds. It includes users with normal or restricted mobility and corresponds to four categories: Task 1 (reading on phone, focus state), Task 2 (closing eyes, relax state), Task 3 (from focus to relax state), and Task 4 (from relax to focus state).

\item {\textbf{EyeState}~\cite{Dua2017}.} This dataset includes 14,980 instances of 14-channel EEG signals recorded by participants performing a task, with a duration of 117 seconds on each subject. The EEG measurements were obtained using the Emotiv EEG Neuroheadset~\cite{Emotiv} and classified into two categories: 1' for eye-closed and 0' for the eye-open state.

\item {\textbf{BCI-2000} \cite{SchalkTBE2004}.} This dataset contains over 1500 one- and two-minute EEG recordings from 109 volunteers. The 64-channel EEG signals from participants can be classified into four tasks: Task 1 (open and close left or right fist), Task 2 (imagine opening and closing left or right fist), Task 3 (open and close both fists or both feet), and Task 4 (imagine opening and closing both fists or both feet).

\item {\textbf{Hybrid}.} This dataset is a combination of the three aforementioned datasets and contains ten categories. The hybrid train data is composed of all the training sets of the three datasets, and consists of 400--800 samples chosen at random from each of the three datasets.
\end{itemize}

\subsection{EEG Features Extraction}
\label{sec:public_dataset}
To show the better performance of our dynamic encoder for feature extraction, we added a linear layer after our dynamic encoder to output classification results, and compared it with the following state-of-the-art baselines:
%
\subsubsection{Baselines}
We implemented the following competitors to validate the substantial advantages of BRIEDGE for EEG classification:
1) \textbf{BiLSTM}~\cite{2005Framewise} can add capacity by forward and backward processing to store and generate more extended range patterns than a single LSTM network.
2) \textbf{EEGNet}~\cite{LawhernJNE2018} is a compact fully convolutional network for EEG-based BCIs.
3) \textbf{Compact-CNN} \cite{WaytowichJNE2018} is a compact ConvNet used for decoding signals from a 12-class SSVEP dataset without the need for user-specific calibration.
4) \textbf{DeepConvNet} and \textbf{ShallowConvNet} \cite{SchirrmeisterHBM2017} are not only novel, promising tools in the EEG decoding toolbox but combined with innovative visualization techniques.
5) \textbf{EEG-TCNet} \cite{IngolfssonSMC2020} is a novel temporal ConvNet, which achieves outstanding accuracy with only a few trainable parameters. 

6) \textbf{DeepBrain} \cite{WuUbiComp2022} is a transformer-base scheme, that eases the construction and utilizes the residual method to enhance the EEG feature  extraction.

\subsubsection{Experimental Setup.}
\label{expsetup}

The configurations of edge server are as follows: CPU: Intel(R) Core i7-7700HQ @2.80GHZ, RAM: 8GB LPDDR4. Our system was implemented with the Pytorch framework and trained on an Nvidia 1060Ti GPU from scratch in a fully-supervised manner. In the training stage, we utilized the Adam Optimizer to compute the stochastic gradient descent. The learning rate of the BRIEDGE was $10^{-3}$, and the weight decay was $10^{-5}$. 
The number of the dynamic encoder output dimension was 128. To avoid missing any temporal correlation in the low-dimensional data, we set some stride of 1 and the kernel size of (1,1) in the brain transformer. In the joint feature extraction, we chose 4 heads for all the multi-head attention mechanism.

 \begin{table}[htb]
    %
	\caption{Feature Extraction Performance with Datasets.}
 \vspace{-3mm}
	\centering
\label{tab_single_datset}\scalebox{0.8}{$
\begin{tabular}{|c|c|c|c|c|c|}
    \hline
    \textbf{Dataset} & \multicolumn{4}{|c|}{Brainlink} 
    \\
    \hline
    Metrics
    & Accuracy & Recall & Precision & F1 Score  \\
    \hline
    BiLSTM & 0.9688 & 0.9688 & 0.9696 & 0.9688 \\ \hline
    EEGNet & 0.9711 & 0.9711 & 0.9713 & 0.9711 \\ \hline
    CompactCNN & 0.9699 & 0.9699 & 0.9702 & 0.9699 \\ \hline
    DeepConvNet & 0.9738 & 0.9738 & 0.9741 & 0.9738 \\ \hline
    ShallowConvNet & 0.7973 & 0.7973 & 0.8652 & 0.7840 \\ \hline
    TCNet & 0.9808 & 0.9808 & 0.9810 & 0.9808 \\ \hline
    DeepBrain &  0.9807 & 0.9738 & 0.9741 & 0.9738 \\ \hline
    \textbf{BRIEDGE} & \textbf{0.9878} & \textbf{0.9821} & \textbf{0.9878} & \textbf{0.9901} \\ \hline
\end{tabular}$}
\scalebox{0.8}{$
\begin{tabular}{|c|c|c|c|c|c|}
    \hline
    \textbf{Dataset} & \multicolumn{4}{|c|}{EyeState} 
    \\
    \hline
    Metrics
    & Accuracy & Recall & Precision & F1 Score  \\
    \hline
    BiLSTM & 0.9101 & 0.9100 & 0.9106 & 0.9099 \\ \hline
    EEGNet &  0.8821 & 0.8650 & 0.8651 & 0.8848\\ \hline
    CompactCNN & 0.7425 & 0.7425 & 0.7425 & 0.7424 \\ \hline
    DeepConvNet & 0.8802 & 0.7475 & 0.7565 & 0.7446 \\ \hline
    ShallowConvNet & 0.6975 & 0.6975 & 0.6992 & 0.6964 \\ \hline
    TCNet & 0.6775 & 0.6775 & 0.6809 & 0.6765 \\ \hline
    DeepBrain & 0.8475 & 0.8475 & 0.8565 & 0.8446 \\ \hline
    \textbf{BRIEDGE} & \textbf{0.9251} & \textbf{0.9150} & \textbf{0.9152} & \textbf{0.9153} \\ \hline
\end{tabular}$}
\scalebox{0.8}{$
\begin{tabular}{|c|c|c|c|c|c|}
    \hline
    \textbf{Dataset} & \multicolumn{4}{|c|}{BCI-2000} 
    \\
    \hline
    Metrics
    & Accuracy & Recall & Precision & F1 Score  \\
    \hline
    BiLSTM & 0.3098 & 0.3098 & 0.3203 & 0.2477\\ \hline
    EEGNet &  0.5986 & 0.2960 & 0.2227 & 0.2465 \\ \hline
    CompactCNN & 0.3656 & 0.3656 & 0.3135 & 0.2877 \\ \hline
    DeepConvNet &  0.5090 & 0.3705 & 0.3896 & 0.3691 \\ \hline
    ShallowConvNet &  0.2539 &  0.2539 &  0.1254 &  0.1644 \\ \hline
    TCNet & 0.5503 & 0.5505 & 0.6004 & 0.5573 \\ \hline
    DeepBrain & 0.5605 & 0.5605 & 0.5696 & 0.5690 \\ \hline
    \textbf{BRIEDGE} & \textbf{0.6602} & \textbf{0.6631} & \textbf{0.6620} & \textbf{0.6651} \\ \hline
\end{tabular}$}
\scalebox{0.8}{$
\begin{tabular}{|c|c|c|c|c|c|}
    \hline
    \textbf{Dataset} & \multicolumn{4}{|c|}{Hybrid} 
    \\
    \hline
    Metrics
    & Accuracy & Recall & Precision & F1 Score  \\
    \hline
    BiLSTM & 0.6372 &  0.6572 &  0.6838 &  0.6623\\ \hline
    EEGNet &  0.6261 &  0.5992 &  0.6018 &  0.5963 \\ \hline
    CompactCNN & 0.6383 &  0.6083 &  0.6164 &  0.6044 \\ \hline
    DeepConvNet &  0.5745 &  0.6045 &  0.6156 &  0.6121 \\ \hline
    ShallowConvNet &  0.5687 &  0.5687 &  0.5821 &  0.5642 \\ \hline
    TCNet & 0.6122 &  0.6122 &  0.6157 &  0.6118 \\ \hline
    DeepBrain & 0.523 &  0.5245 &  0.515 &  0.5231 \\ \hline
    \textbf{BRIEDGE} & \textbf{0.8650} &  \textbf{0.8591} &  \textbf{0.8648} &  \textbf{0.8651} \\ \hline
\end{tabular}$}
\end{table}
\vspace{-2mm}
\subsubsection{Single Dataset Classification}
\begin{figure*}[tb]
	\begin{center}
		\begin{tabular}{ccccc}
			\includegraphics[width=0.38\columnwidth]{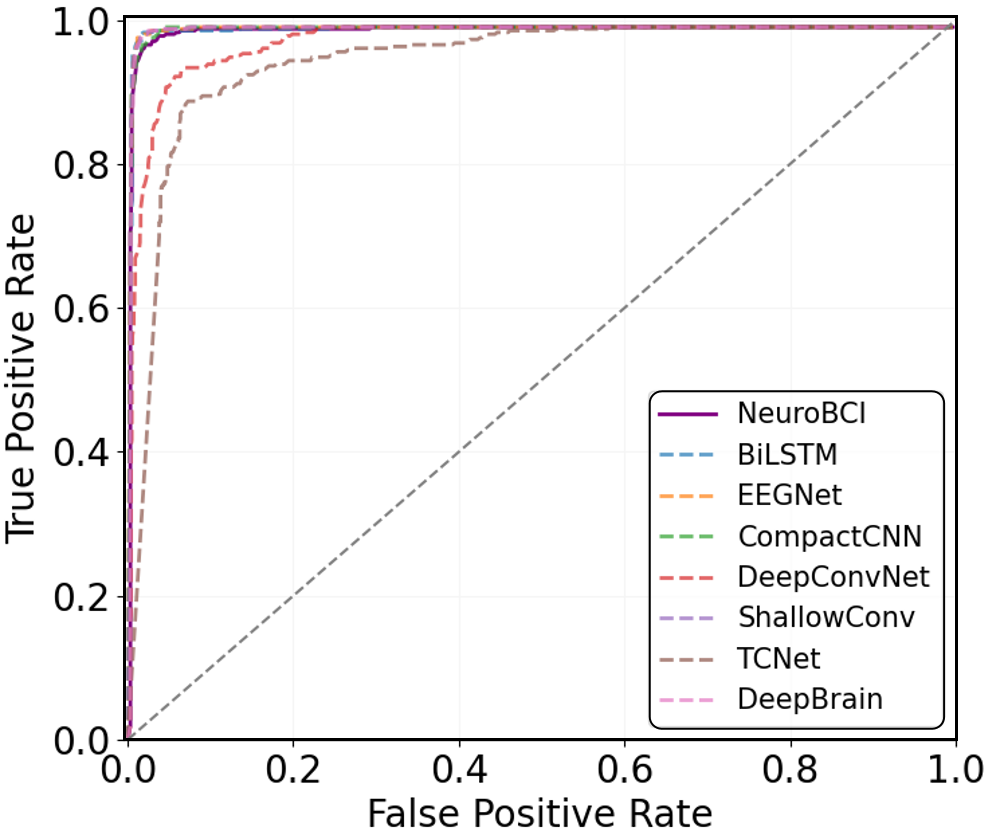}&
			\includegraphics[width=0.38\columnwidth]{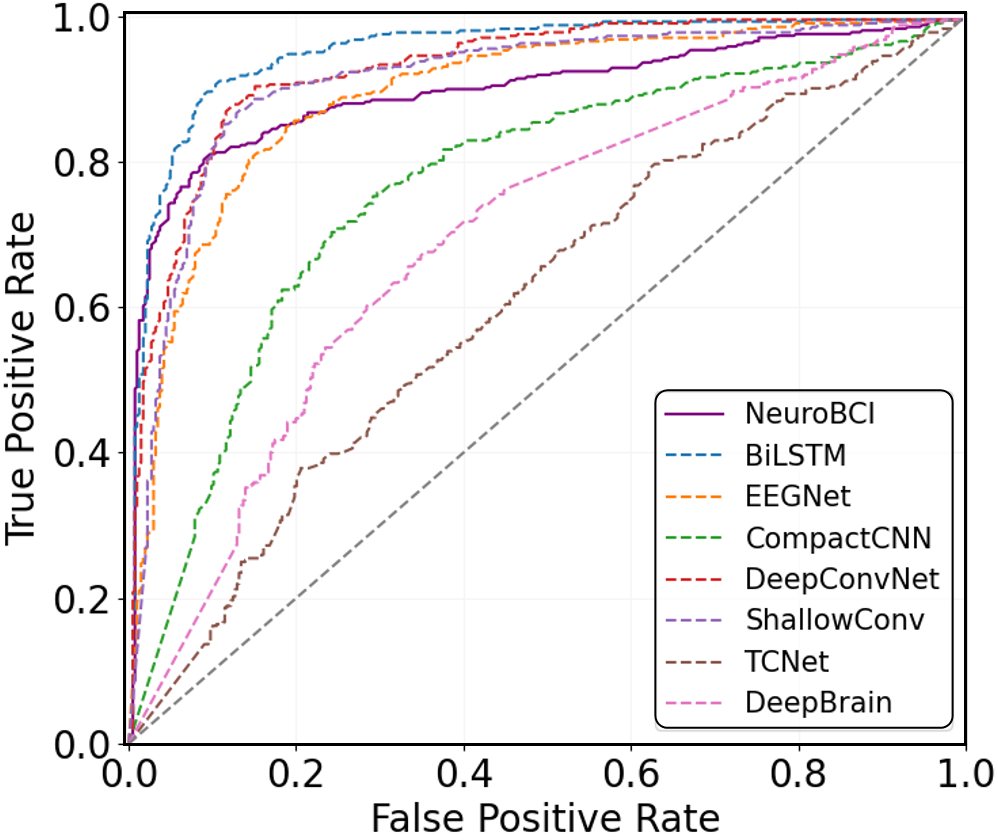}& \includegraphics[width=0.38\columnwidth]{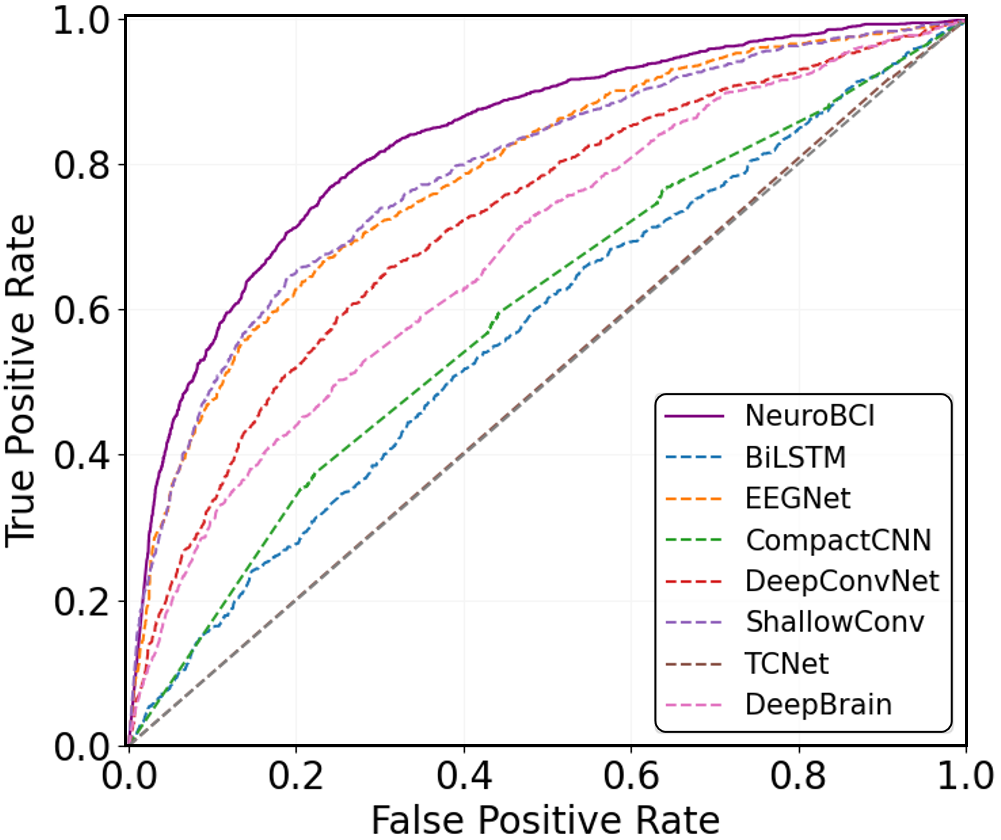}&\includegraphics[width=0.38\columnwidth]{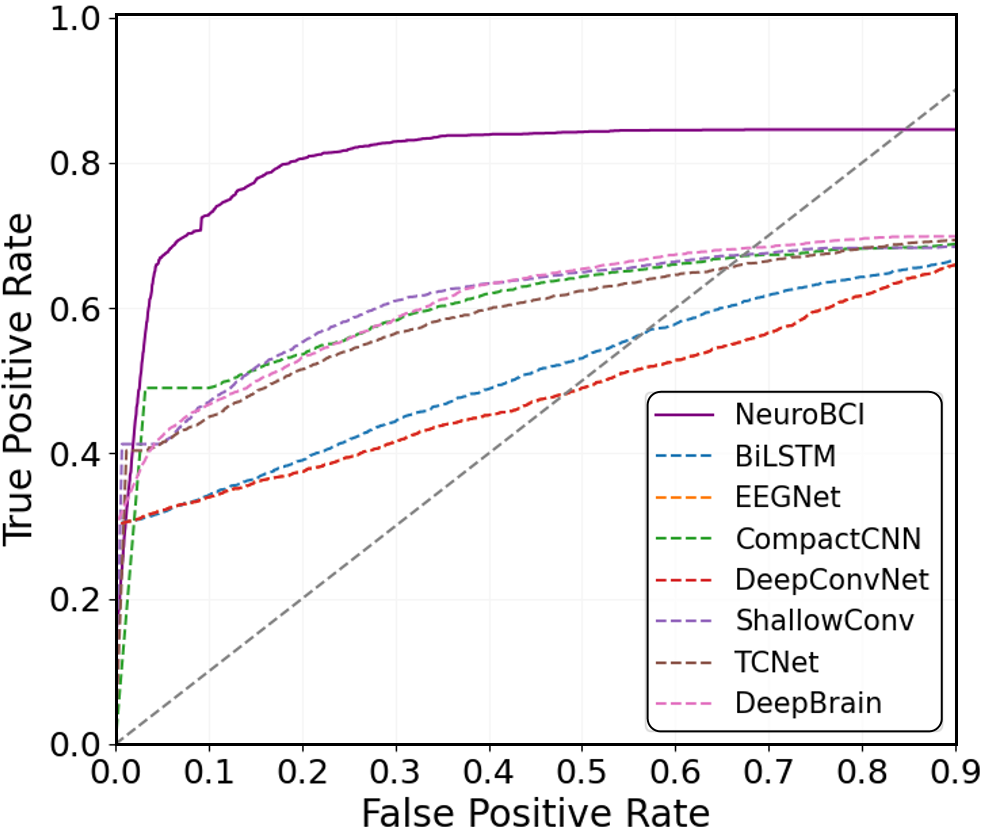}&
			\includegraphics[width=0.37\columnwidth]{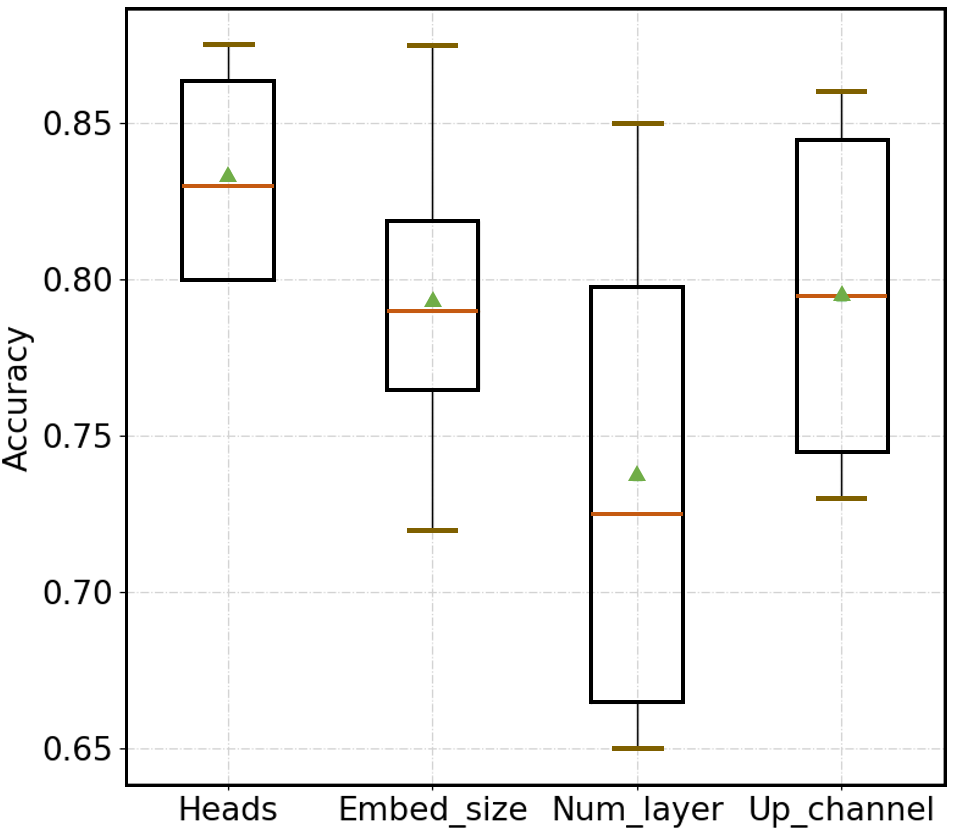}\\
			{\scriptsize (a) Brainlink}&
			{\scriptsize (b) EyeState}&
			{\scriptsize (c) BCI-2000}&      {\scriptsize (d) Hybrid} &
			{\scriptsize (e) Parameter Sensitivity} \\
		\end{tabular}
	\end{center}
	\vspace{-3mm}
	\caption{\textcolor{black}{The performance of ROC and parameter sensitivity on heterogeneous datasets.}}
	\label{fig:ROC}
 \vspace{-3mm}
\end{figure*}

\ 
We first conduct the experiments using three single datasets (Brainlink, EyeState, and BCI-2000). Three single datasets here correspond to the data acquisition from three specific EEG devices with diverse amounts of electrodes. The dynamic process in our encoder is designed for multiple forms of EEG data, but it also works well when the input data only has a single form.
For each method, we split 80\% from every single dataset as the training dataset and 20\% as the testing dataset, respectively.  Each method runs several rounds till convergence using the same hyperparameter settings and training dataset. The same testing dataset is then used to evaluate each method. As depicted in Table~\ref{tab_single_datset}, we have compared some EEG-based and image-based neural networks, which have made outstanding contributions to their field, and our method outperforms them in the classification of these three datasets. The result shows that our method has a better performance when just inputting a single type of EEG data, which indicates that our system could be applied to the scenario with just one EEG device. Here, we utilize ROC Curve (Receiver Operating Characteristic Curve) to directly present the classification performance. The higher the ROC curve, the better the classification. As shown in Fig.~\ref{fig:ROC}, BRIEDGE shows the best classification performance in an experiment with the Brainlink dataset while performing well in the experiment with the EyeState dataset. It can be also observed that BRIEDGE is able to learn the high-dimensional data well in the experiment with the BCI-2000 dataset. Compared with other methods like TCNet and DeepBrain, BRIEDGE has a better performance in learning different types of EEG data.
\vspace{-2mm}

\subsubsection{Hybrid Dataset Classification} To verify the performance of BRIEDGE on multi-typed EEG data, we integrate the above mentioned three datasets as a hybrid dataset. Since  EEG data collected by different BCI devices have different dimensions, we use an embedding layer to map heterogeneous data into the latent space of the same dimension.  
 
As shown in Table~\ref{tab_single_datset}, BRIEDGE outperforms the other baselines in the hybrid dataset, which is more complex than a single dataset, indicating that dynamic feature extraction can be a reasonable learning method in the scenario of multiple types of EEG data flow. We also observe that the accuracy of the hybrid dataset is not simply averaging the accuracy of the previous three datasets because different data types could confuse the training of neural networks. 

In addition, we use the ROC metric and parameter sensitivity analysis to demonstrate the stability of each model during the inference classification process, as shown in Fig.~\ref{fig:ROC}. The ROC performance with four datasets are presented in Fig.~\ref{fig:ROC} (a)-(d). The result shows that BRIEDGE is the most stable method in the classification predicting procedure. The parameter sensitivity experiment in Fig.~\ref{fig:ROC} (e) can find sensitive parameters in BRIEDGE for the hybrid dataset. Four sensitive parameters are the heads of self-attention mechanism in joint feature extraction (Heads), the embed size of dynamic encoder (Embed\_size), the stack number of brain transformer (Num\_layer), and the up-dimension channel of AvgpoolingEEGNet (Up\_channel) respectively in Fig.~\ref{fig:ROC} (e).

\subsection{\textcolor{black}{Encoding-decoding Communication}}
In this section, we compare the performance of BRIEDGE, the semi-traditional system with an extra feature coding for EEG transmission under the additive white Gaussian noise (AWGN) channels, where the accurate channel state information (CSI) is assumed. 
\subsubsection{Baselines}\label{C-E}

To verify the semantic encoding and decoding capability and the channel encoding and decoding capability of BRIEDGE, we use the following five baselines: 
1) \textbf{BRIEDGE-Lite}.
The semantic encoder and semantic decoder in BRIEDGE-Lite have the same structure as in BRIEDGE, except that the channel encoder and channel decoder each have only one fully-connected layer.

2) \textbf{BRIEDGE-RS}. 
This is the conventional Reed-Solomon (RS) coding~\cite{reedSIAM1960} for channel coding, with the same semantic coding as BRIEDGE.
3) \textbf{EEGNet-SC}. 
It is the EEGNet-based encoding-decoding communication framework, comprising of the EEGNet-based semantic encoder, and decoder, with the rest of the structure identical to that of BRIEDGE.

4) \textbf{EEGNet-SC-Lite}. 
This is based on EEGNet-SC with a fully-connected layer being used in the channel encoder and the channel decoder, respectively, and the rest of the structure is the same as that in EEGNet-SC.
5) \textbf{EEGNet-SC-RS}. It uses the standard Reed-Solomon (RS) coding for channel coding, and the remaining structure is the same as EEGNet-SC.

\subsubsection{Varying Channel Environments and Datasets}

\begin{figure}[htb]
\begin{center}
\begin{tabular}{cc}
\includegraphics[width=3.6cm]{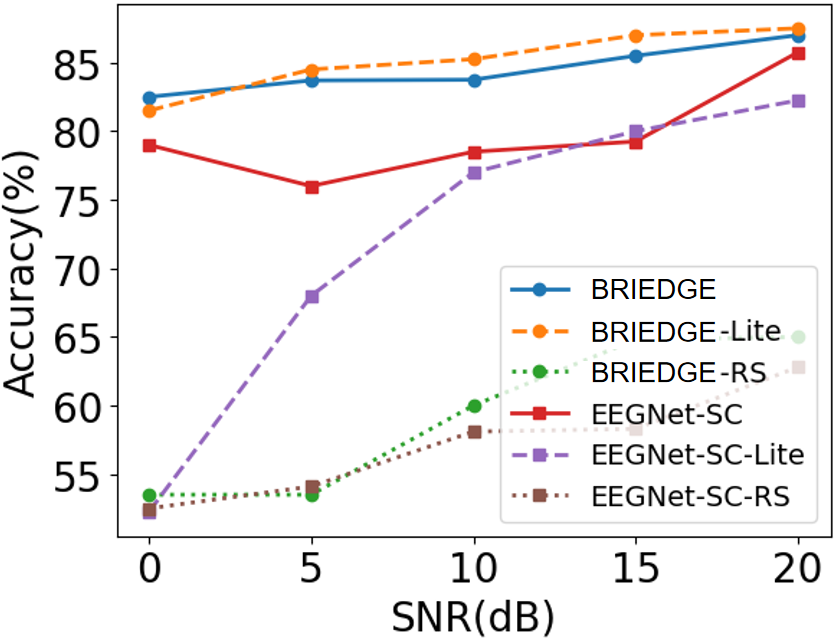} & \includegraphics[width=3.6cm]{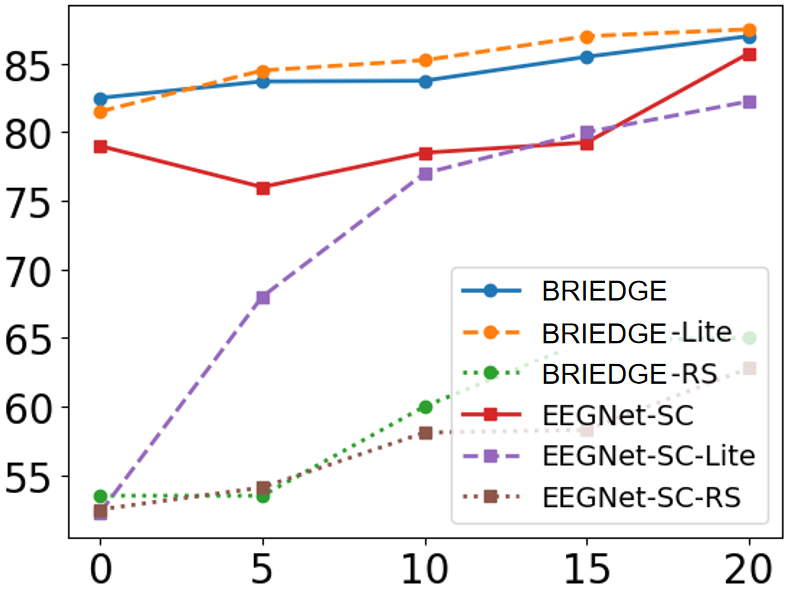} 
\\ 
{\scriptsize(a) Brainlink} & {\scriptsize (b) EyeState} \\
\includegraphics[width=3.6cm]{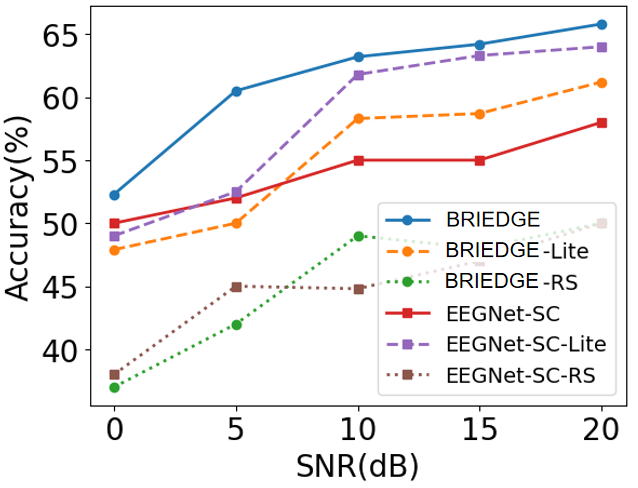} & \includegraphics[width=3.6cm]{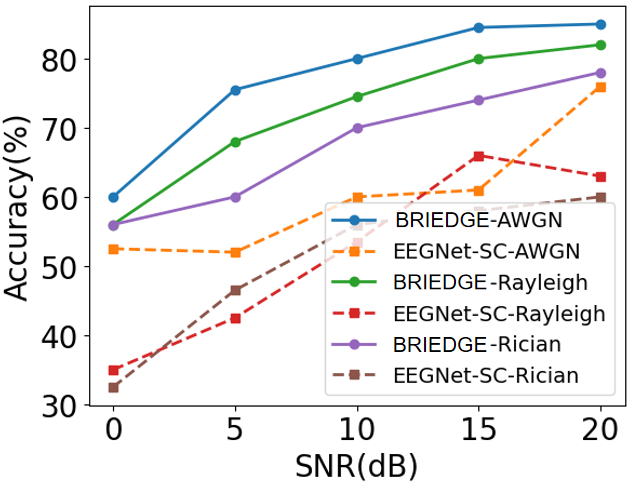} \\ {\scriptsize(c) BCI-2000} & {\scriptsize (d) Hybrid} \\[5pt]
\end{tabular}
\vspace{-5mm}
\caption{Performance of encoding-decoding communication in environments with various signal-to-noise ratio (SNR).}
\label{fig:sc}
\end{center}
\end{figure}

Fig.~\ref{fig:sc} (a)-(c) compare the transmission accuracy of BRIEDGE and EEGNet-SC based on their variants on three different datasets and different SNR communication environments under AWGN channel. We conclude two important conclusions from the experiment. Firstly, we can observe that  EEGNet-SC-RS and BRIEDGE-RS based on traditional channel coding are significantly weaker than the other methods. This indicates that our BRIEDGE model on each dataset can deliver state-of-the-art transmission accuracy compared with the semi-traditional communication framework. This may be because RS coding is linear block coding with a long block length, and can correct a long series of bits, however, BRIEDGE and other neural network-based encoding-decoding communication methods are not only suitable for short block length but also perform better in decoding long block-length. Secondly, when SNR is low, BRIEDGE, BRIEDGE-Lite, and EEGNet-SC achieve higher transmission accuracy, which indicates that the performances of BRIEDGE and EEGNet are quite stable and effective even under a poor communication environment. Indeed, they perform consistently well when SNR ranges from 0 to 20. The stability of BRIEDGE is due to the superior extraction and recovery of EEG signals in the process of semantic coding and decoding.
To highlight the merit of BRIEDGE over EEGNet, we further evaluate the transmission accuracy on the hybrid dataset under different SNR based on BRIEDGE and EEGNet in AWGN channel environments.

\vspace{-3mm}

\section{Interaction Evaluation}
As demonstrated in Section~\ref{sec:expriment1}, our system exhibits superior classification and communication performance, indicating its ability to effectively address the challenges of heterogeneous input data and communication loss, while maintaining robust data security. In this section, we focus on the evaluation of real-time systems to ensure that our system is acceptable to users. \textcolor{black}{We have invited 24 participants to our evaluation, and half of them participated in both the data collection and interaction stages (PP subjects), while the rest just participated in the latter stage (NP subjects).} As for the data collection and user study, we made a statement on the rights and interests of participants in the questionnaire. Participants who responded to the questionnaire stated that they had participated voluntarily, were over 18 years old, were aware of the publication of research results, and confirmed that their data were collected anonymously. With their consent, we recorded the data for further analysis. Moreover, they are aware that they can quit the experiment and request to remove their data at any time during or after the user study.

\subsection{Data Collection and Usage}
\vspace{-3mm}
\begin{figure}[htb]
\begin{center}
\begin{tabular}{ccc}
\includegraphics[width=2.5cm]{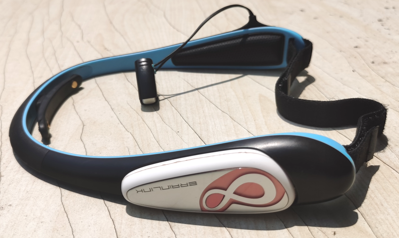} & \includegraphics[width=2.5cm]{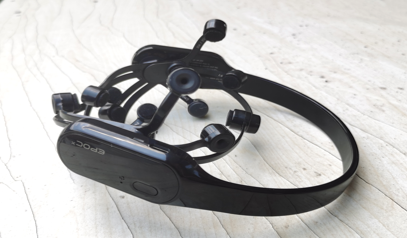} & \includegraphics[width=2.5cm]{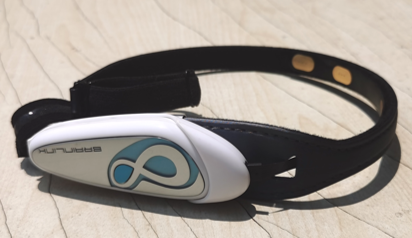} 
\\ 
{\scriptsize(a) Brainlink pro.} & {\scriptsize (b) Emotiv Epoc.} & {\scriptsize(c) Brainlink lite.} 
\end{tabular}
\\
\begin{tabular}{c|c}
\includegraphics[width=6.5cm]{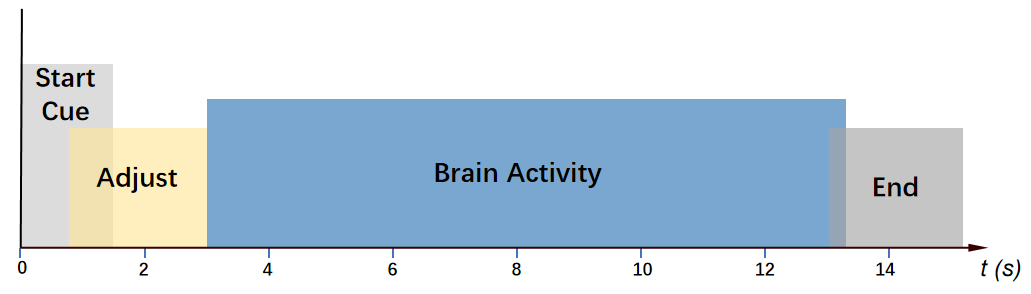}
\\
{\scriptsize(d) Data collection session procedure.}
\end{tabular}

\vspace{-4mm}
\caption{Devices in interaction evaluation.}
\label{fig:device}
\end{center}
\end{figure}
\vspace{-3mm}
\textbf{\textcolor{black}{Data Collection and Pre-trained Dataset.}} 
\textcolor{black}{PP subjects were engaged in the data collection stage for BRIEDGE pre-training. Collecting 70–80 samples (15–20 samples per category) from each subject using every EEG device, illustrated in Fig.~\ref{fig:device} (a) (b) (c), to form evaluation datasets. PP subjects followed a session paradigm shown in Figure~\ref{fig:device} (d). Data samples were pre-processed by excluding the first 3 seconds and the last 2 seconds of each sample. Then, three pre-train datasets were established: PRO Dataset (including 4 categories of relax-focus state), Emotiv Dataset (including 8 categories: 4 categories of relax-focus state and 4 categories of four-direction motor imagination), and LITE Dataset (including 4 categories of relax-focus state). 80\% of each dataset is designated as the training set, while the remaining 20\% is used for testing and performance evaluation.}

\textbf{\textcolor{black}{System Pre-training Setting.}} 
\textcolor{black}{
We hybridized the mentioned three training datasets (from 3 kinds of devices) and trained BRIEDGE, to study heterogeneous data learning performance and user experience evaluation. The data was randomly fed into the system without considering its size. In terms of the parameter setting, the batch size was set to 32, the learning rate to 0.0005, and the training epoch to 50.
After training with the three-hybridized training dataset, our system achieved classification accuracy of 93.50\%, 95.46\%, and 97.33\% on testing datasets from three devices, respectively. We subsequently saved the parameters of the semantic encoder, channel encoder, and semantic decoder \& channel decoder as SE model, CE model, and SD-CD model, respectively.}

\subsection{Systematical Performance and User Experience}
In this section, we evaluate the performance of our pre-trained system and address realistic delay issues, followed by a performance analysis and user experience experiments. Our system includes an edge server to host computing resources for EEG signals and forward recognized control commands to three robots: a wheel robot, a mechanical arm, and an aerial drone. We used Bluetooth as the physical channel between the transmitter and receiver, where a transmitter can be a mobile server or portable server (NVIDIA JETSON XAVIER NX DEVELOPER KIT). The server (with SE model \& CE model) and the robots (with SD-CD model) exchange command information through Bluetooth.
\begin{figure*}[htb]
\begin{center}
\begin{tabular}{cccc}
\includegraphics[width=3.4cm]{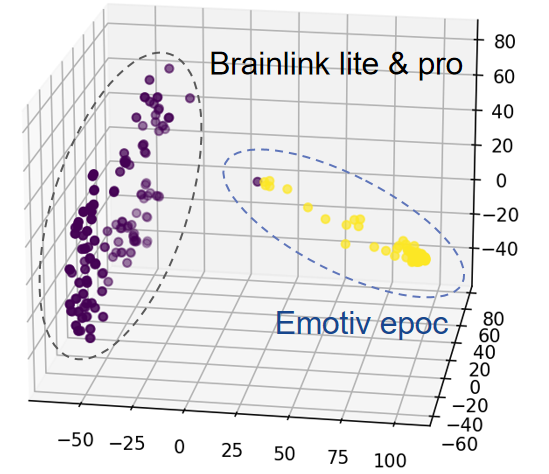} & \includegraphics[width=3.7cm]{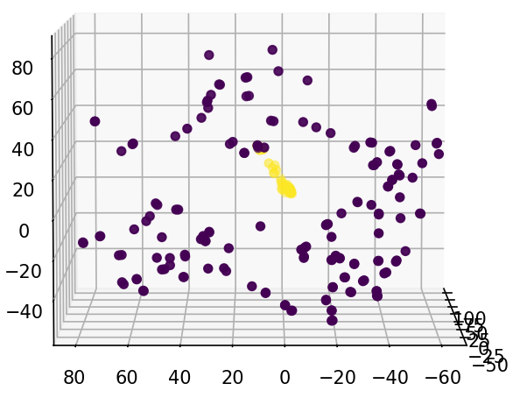} & \includegraphics[width=3.7cm]{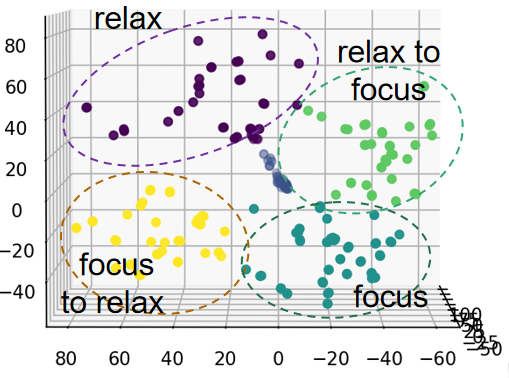} & 
\includegraphics[width=3.4cm]{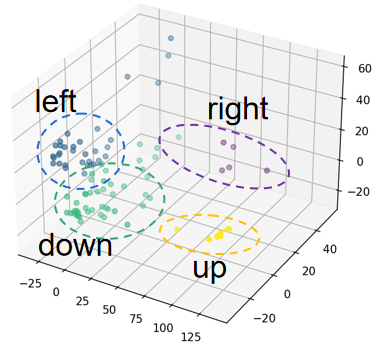}
\\ 
{\scriptsize (a) Clusters of devices} & {\scriptsize (b) Another perspective} & {\scriptsize (c) Clusters of Brainlink commands} & {\scriptsize (d) Clusters of Emotiv commands} \\
\end{tabular}
\vspace{-5mm}
\caption{Joint features cluster distribution of three-mixed datasets.}
\label{fig:first_joint}
\end{center}
\end{figure*}

\begin{figure*}[htb]
\begin{center}
\begin{tabular}{cccc}
\includegraphics[width=3.6cm]{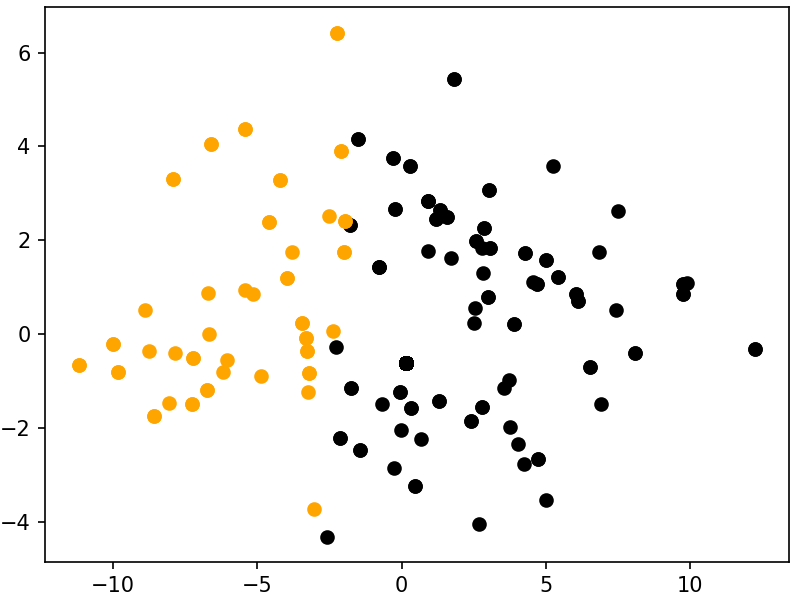} & \includegraphics[width=3.4cm]{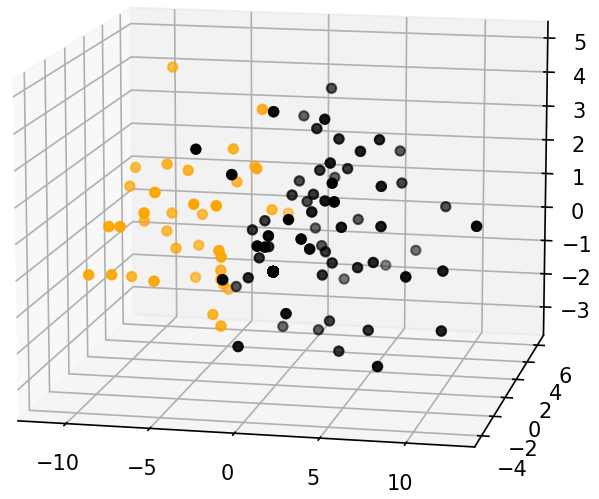} & \includegraphics[width=3.6cm]{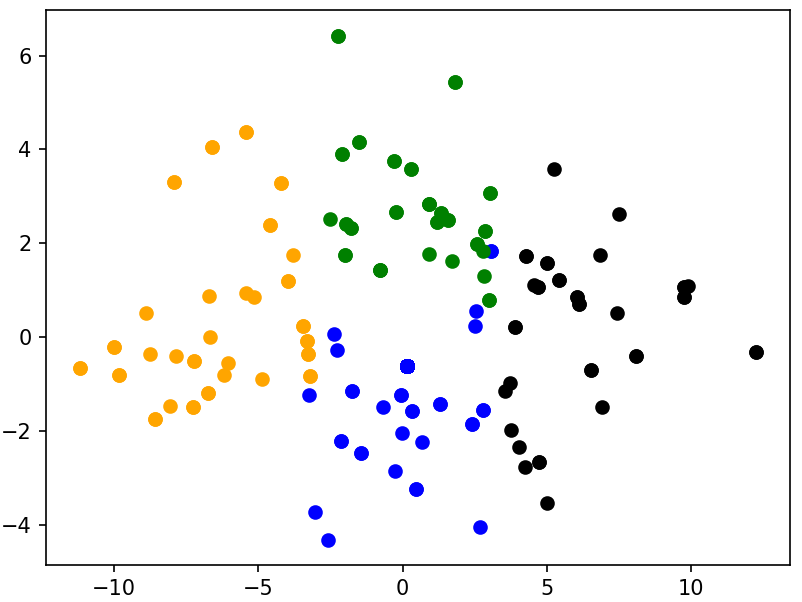} & \includegraphics[width=3.4cm]{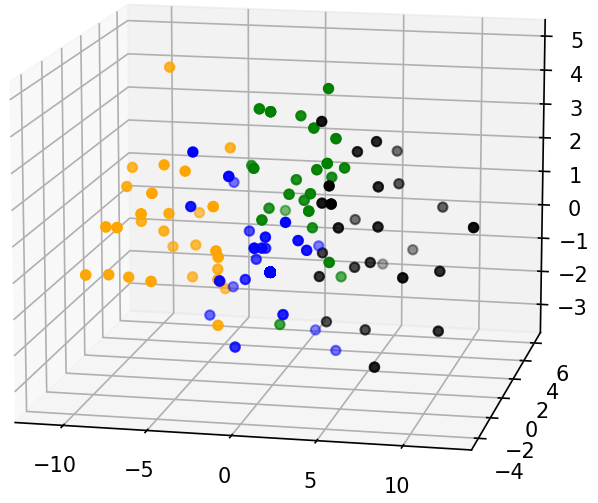}
\\ 
{\scriptsize (a) 2-group clusters in 2D } & {\scriptsize (b) 2-group clusters in 3D}  & {\scriptsize (c) 4-group clusters in 2D } & {\scriptsize (d) 4-group clusters in 3D} \\
\end{tabular}
\vspace{-5mm}
\caption{Joint features cluster distribution of four-mixed datasets.}
\label{fig:second_joint}
\end{center}
\end{figure*}
\subsubsection{Systematical Performance and Analysis}
\textcolor{black}{Firstly, we extracted joint features from the joint features extraction module to analyze. All data types were mapped into a latent space of [128, 128]. PCA and Kmeans mechanisms are used to visualize the clusters of the testing dataset mixed with the Emotiv Motor (4 categories of four-direction motor imagination), PRO, and LITE testing datasets. As demonstrated in Fig.~\ref{fig:first_joint} (a) and (b), The notable difference between Emotiv EPOC device data, and Brainlink devices. The successful separation of specific tasks is in Fig.~\ref{fig:first_joint} (c) and (d). Results validate BRIEDGE can recognize high-dimensional and low-dimensional data along diverse processing paths.}

\textcolor{black}{Subsequently, we merged all datasets with all categories to assess our system, visualizing the 2-group clusters of joint features in Fig.~\ref{fig:second_joint} (a) (b), where the orange points denote Brainlink devices and the black points represent Emotiv EPOC. The 4-group clusters are also presented in Fig.~\ref{fig:second_joint} (c) (d), where 4 types of data can be separated. The data from similar devices or with similar tasks is closer; {\em e.g.} the green and black dots come from Emotiv datasets of two types of brain tasks; the orange and blue dots come from LITE and PRO, respectively. The results signify that BRIDGE retains the data differences while retaining the data similarities, enabling the model to classify devices while classifying specific EEG signals. It proves that BRIEDGE can address the issue of the great gap between heterogeneous data mentioned in Section~\ref{sec:motivation}.}
\begin{figure*}[htb]
    \centering
    \includegraphics[width=1\linewidth]{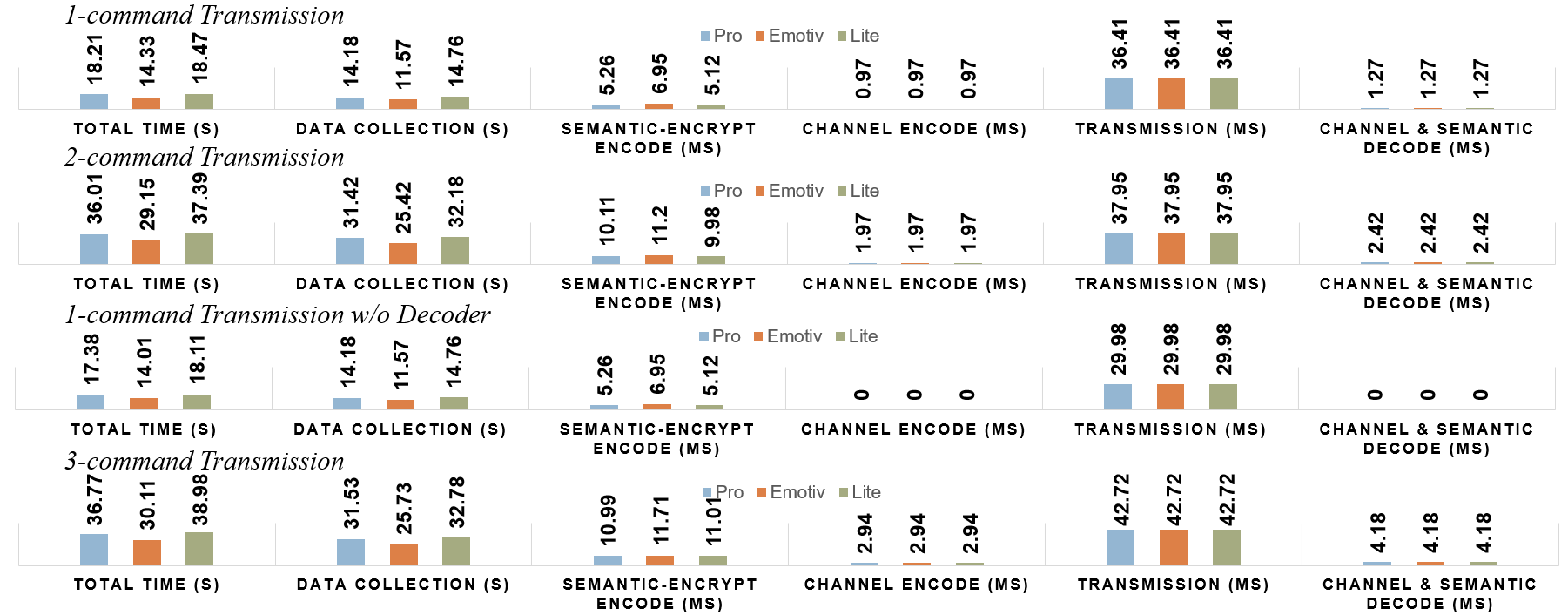}
    \vspace{-5mm}
    \caption {\textcolor{black}{Command transmission time cost on the device-to-server communication.}}
    \label{fig:communication_result} 
\end{figure*}
\textcolor{black}{Furthermore, we evaluated control communication time costs. Four experiments were designed to showcase communication delays: transmit one or two commands (1-Command or 2-Command EEG-based Semantic), transmit one command with a traditional process (1-Command Tradition), and three devices transmit commands simultaneously for control (3-Device Real-time Control). Results in Fig.~\ref{fig:communication_result} indicate data collection comprises approximately 80\% of time costs; however, collection time is related to user brain activity duration. This demonstrates our encoding-decoding communication process ensures accuracy while maintaining brain-to-robot transmission smoothness.}

\subsubsection{\textcolor{black}{User Experience Study}}
\begin{figure*}[htb]
\begin{center}
\begin{tabular}{cccc}
\includegraphics[width=4cm]{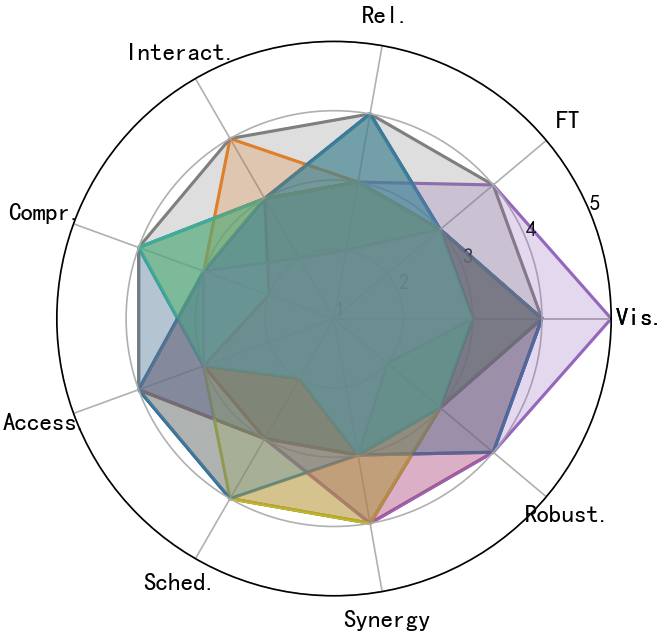} & \includegraphics[width=4cm]{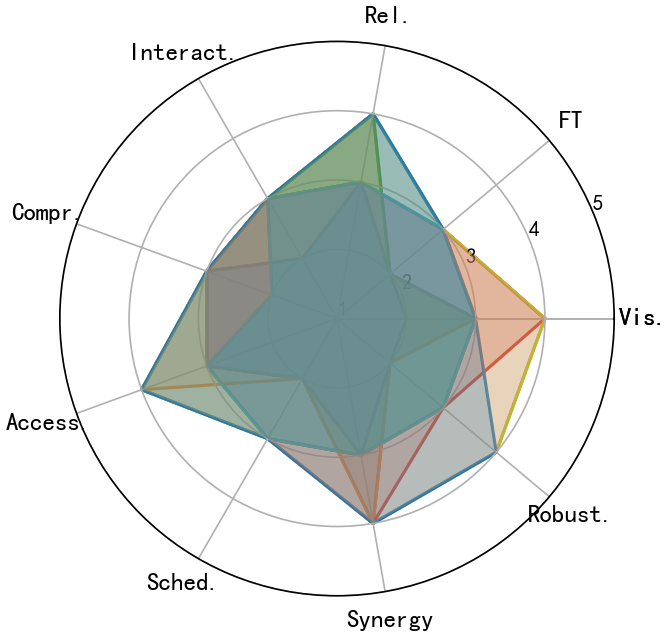} 
\\ 
{\scriptsize \textcolor{black}{(a) The subjects previously participated (PP) in data collection.}} & {\scriptsize \textcolor{black}{(b) The subjects not participated (NP) in data collection.}} \\[5pt]
\end{tabular}
\vspace{-4mm}
\caption{\textcolor{black}{Subjects' feedback. \scriptsize{The collected data from PP subjects is fed into the BRIEDGE for training, and vice versa for NP subjects.}}}
\label{fig:user_study}
\end{center}
\end{figure*}

\textcolor{black}{In this section, we invited subjects to interact with BRIEDGE, controlling specific robots or collaborating. Subjects are divided into two groups: previously participated in the data collection stage (PP subjects) or not participated in the stage (NP subjects), to find out whether the pre-training process impacts the user experience. During the experiments, we invited subjects to control specific robots using three EEG devices, and then every three subjects collaborated through BRIEDGE to conduct a task, transferring an object from one place to another. After all the interaction experiments, subjects filled in a questionnaire on user experience. Partial results in Fig.~\ref{fig:user_study} show that the PP subjects are more positive and willing to score higher than the NP subjects. This suggests that the data collection phase and pre-training process play a significant role in shaping the performance of BRIEDGE, posing a challenge for personalized system design. Here, the abbreviations Vis., FT, Rel., Interact., Compr., Access., Sched., Synergy, and Robust. correspond to the following concepts: Visibility, Fault Tolerance, Relevance, Interactivity, Comprehensibility, Accessibility, Schedule, Synergy, and Robustness, respectively. Details of the questionnaire and subjects' feedback can be found in the supplementary materials.}
\subsection{Model Compression for Edge Portable Server}

We deploy DJI Manifold 2-G~\cite{Manifold2} as a mobile and portable box as the edge portable server to test edge AI for interaction. This edge portable server is configured with the following embedded computing resources: NVIDIA Pascal GPU (256-core), NVIDIA Denver 2 (Dual-Core) and ARM Cortext-A57 CPU (Quad-Core), 8GB LPDDR4 RAM, 32GB eMMC ROM and 128GB SATA-SSD. To ensure the performance on the resource-constrained edge portable server, we implement model compression methods as described in Section~\ref{compression} to receive EEG signals, encode EEG signals, classify EEG signals, and transmit EEG signals. 

\textcolor{black}{Linear layers and convolutional layers will be mainly processed. Firstly, we calculated the standard deviations of the four weight matrices multiplied by $S=1.5$ for convolutional layers and $S=0.25$  for linear layers to get thresholds below which the weight parameters are set to 0 using the weight pruning method. Subsequently, we clustered the weight matrices of different layers into $2^4$ groups using the weight-sharing method. Finally, we utilized the half-quantization method to convert the storage type from float 32 to float 16. After the compression, the trained model is saved in ONNX format~\cite{bai2019onnx} for fast and stable performance on embedded devices, where onnx\_sparsify script is used further to compress the sparse parameters after the above compression.}

\begin{figure}[htb]
	\begin{center}
		\begin{tabular}{cc}
			\includegraphics[width=0.48\columnwidth]{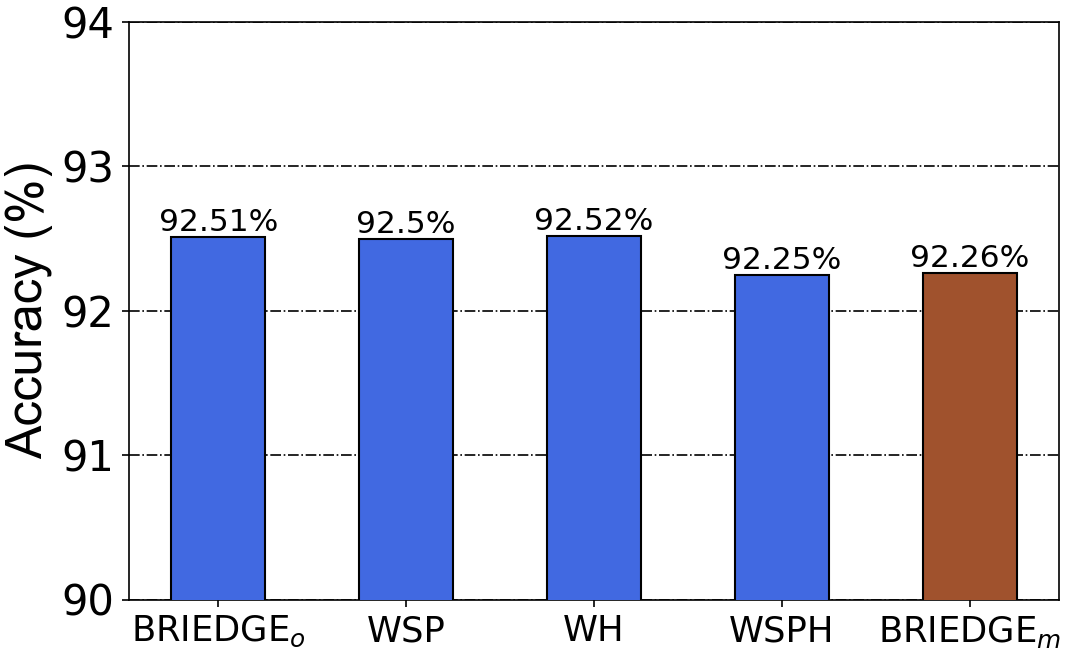}&
			\includegraphics[width=0.48\columnwidth]{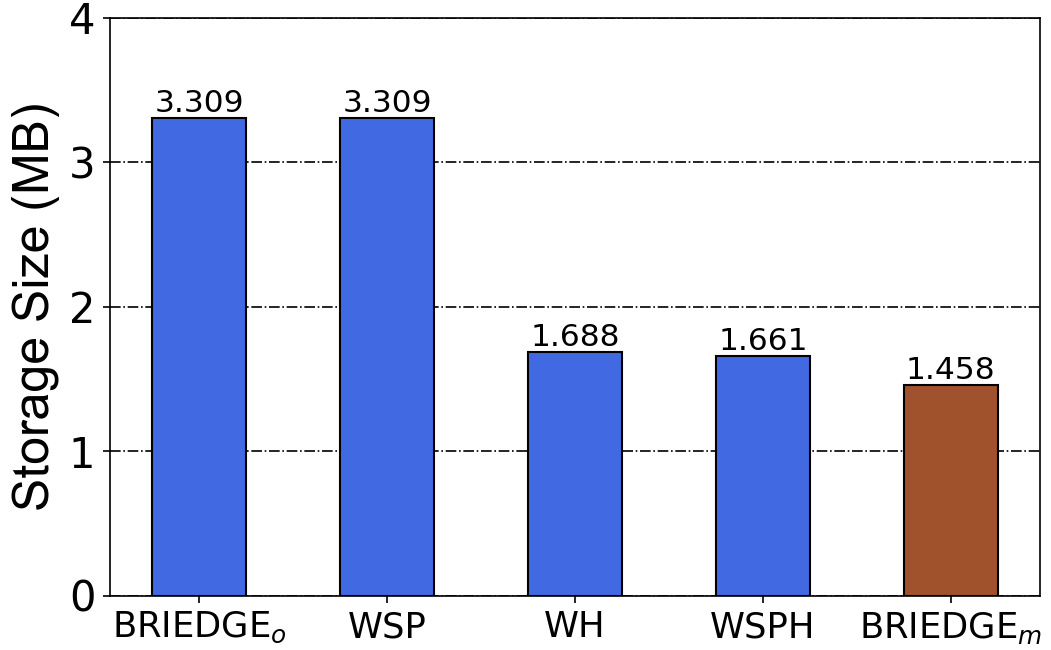}\\
			{\scriptsize (a) Accuracy on Eyestate dataset}& {\scriptsize (b) Storage size}
		\end{tabular}
  \\
  \begin{tabular}{c}
       \includegraphics[width=1\columnwidth]{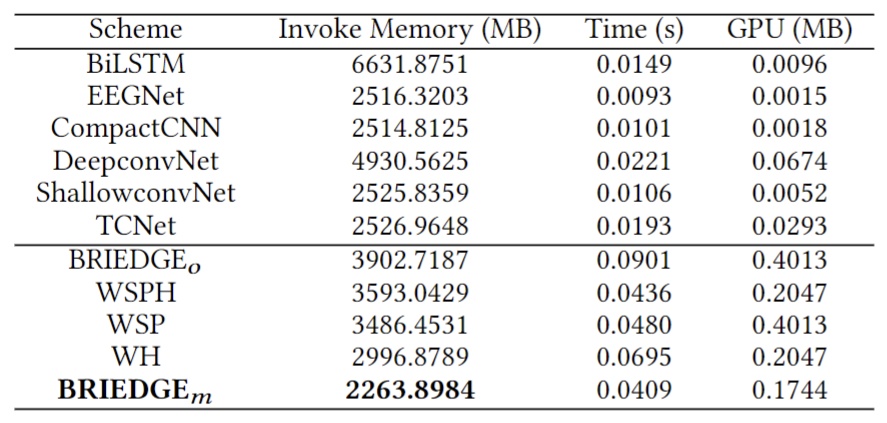}
       \\
       {\scriptsize (c) Performance comparison on edge portable server} 
  \end{tabular}
	\end{center}
	\vspace{-3mm}
	\caption{The performance after the model compression.}
	\label{fig:compression}
\end{figure}

To test the performance on multi-task scenarios, we run experiments on the above-mentioned hybrid dataset in Section~\ref{sec:public_dataset}. Through the model compression, the size of the saved model is reduced significantly from 3.309MB to 1.458MB, whereas the accuracy only decreases by 0.25\% at the same time. The detailed statistics are shown in Fig.~\ref{fig:compression}. The BRIEDGE after our model compression method is named as \textbf{BRIEDGE$_m$}. We compared it with different compression methods including the combination of weight sharing and weight pruning (namely \textbf{WSP}), weight halving (namely \textbf{WH}), the combination of weight sharing, weight pruning and weight halving (namely \textbf{WSPH}), and the original BRIEDGE without any compression methods (namely \textbf{BRIEDGE$_o$}), respectively.

To demonstrate the enhanced performance resulting from model compression, we conducted additional comparisons with BiLSTM, DeepconvNet, EEGNet, CompactCNN, ShallowconvNet, and TCNet using the hybrid dataset on the edge portable server. The experimental results presented in Fig.~\ref{fig:compression} (c) verify that the BRIEDGE$_m$ running on a resource-constrained edge portable server can guarantee efficient EEG signal classification and encoding-decoding communication. Meanwhile, It is observed that our BRIEDGE$_m$ consumes less cost on invoke memory than others, which demonstrates that BRIEDGE$_m$ fits the multi-task for multi-robot scenarios despite limited computing resources. It also proves that our BRIEDGE has the potential to meet the practical requirements of BCI applications.
\textcolor{black}{\section{Discussion and Envision}}
\subsection{Discussion}
\subsubsection{\textcolor{black}{Interactive Scheme Designed to Provide a New Method of Daily Collaborating Tasks}} \textcolor{black}{Individuals encountering daily movement challenges often require assistive devices or help from others to perform various physical tasks. Yet, research suggests that many simple tasks significantly restrict users' daily lives, and most tasks can be assisted by one or more robots. Our system provides a new interactive scheme for multiple users and multiple robots using BCI techniques. Users can collaboratively choose different robots and complete tasks with other collaborators. In light of the HCI community's heightened interest in understanding and supporting vulnerable groups, we believe our research marks the initiation of current BCI development from a people-centered perspective.}

\subsubsection{\textcolor{black}{Using Encoder-Decoder Framework to Enhance BCI with Heterogeneous Data}} \textcolor{black}{Using BCI to assist users' daily lives faces challenges. Various data conflicts appear in a BCI system addressing multiple user scenarios. Designing a designated algorithm for each civil EEG device increases developmental difficulties, halting the general use of BCI. We used an encoder-decoder framework to design an adaptable BCI system for diverse EEG devices, significantly increasing the possibility of providing an accessible and performance-guaranteed BCI to most task collaborators and communities of people with mobility difficulties. Moreover, BCI is designed to be used at any time. Therefore, we believe that our work can accelerate the options of BCIs for improving the lives of people.}

\subsubsection{Generality of BRIEDGE} Though we evaluated three datasets and a hybrid dataset as representative EEG data, BRIEDGE is designed to support more types of heterogeneous data corresponding to various EEG collection devices. It also can be generalized to support multimodal fusion of diverse interactive physiological signals. In this sense, BRIEDGE is a generic system for achieving the multi-user multi-agent interaction with complex input data. Meanwhile, our system is capable of handling data proliferation efficiently because its computational time cost scales linearly with the increase in the number of commands.

\subsubsection{\textcolor{black}{Implications for Future BCI}}\textcolor{black}{In the field of human-computer interaction, notable efforts have focused on exploring the applications and advancement of BCI. Despite this, BCI is yet to be recognized as a technology suitable for widespread use in everyday life. Typically characterized by high thresholds, advanced technology, and high costs, BCI shares similarities with the early stages of gesture and voice interaction. Therefore, it is valuable to think about what features and aspects of the BCI system are essential in assisting specific groups of people with unique needs or characteristics.}
\textcolor{black}{This research proposes BRIEDGE, representing the first system that supports multiple EEG devices to interact with multiple agents. Our system can efficiently capture the different types of EEG data. This is achieved by encoding EEG features dynamically meanwhile extracting the joint features for a deeper understanding of the temporal and spatial dependency. The proposed systematic method for addressing the heterogeneous bioelectric signals can be applied for systems based on physiological electrical signals such as muscle signals. Further, multimodal bioelectric signal fusing can be integrated into our system which is a potential aspect. We envision BRIEDGE could become a useful system for future multi-to-multi BCI applications.}

\subsection{Limitation and Future Work}
\subsubsection{Limitations} 

\textbf{Deployment on Edge Portable Server.} \textcolor{black}{Current BRIEDGE has achieved a higher classification accuracy of EEG commands in a noisy transmission environment than existing approaches, and model compression reduces the computational cost for deployment on an edge portable server. However, its computational cost has room to be reduced for resource-constrained edge portable servers due to the dynamic encoding process involved in the system. We will optimize the algorithm structure and deploy measurements to reduce the cost in the future.} 

\textbf{Latency in Real-life Testing.} To reduce collection time costs, we're working on enhancing the accuracy of short-term EEG data classification. However, the challenge lies in the limitations of low-cost EEG devices, which may not match the signal quality of pricier alternatives. Despite optimizing Brainlink devices for 15-second tasks, there's room for improvement in short-term accuracy. We aim to balance system time and accuracy while considering hardware constraints. We believe refining our algorithm and exploring new methods can enhance short-term EEG data classification efficiency.

\textbf{Personality.} \textcolor{black}{The user experience study highlights the importance of addressing cross-subject and personalized issues. Therefore, we plan to incorporate more adaptive approaches to address the cross-subject requirement, such as transfer learning or domain adaptation. By doing so, we will develop a more robust and personalized brain-robot interface system that can adapt to individual differences in brain signals and improve user experience.}

\subsubsection{Future Work}
Initially, we will enhance our model by incorporating state-of-the-art data processing techniques, including contrastive learning and transfer learning, to tackle the challenges posed by large-scale heterogeneous datasets. Simultaneously, a signal-to-command mechanism that prioritizes simplicity will be developed, potentially within a more adaptive encoding-decoding communication framework. Additionally, it is imperative to address individual preferences, encompassing customized control habits and personalized EEG signal representations, especially in scenarios involving multiple users and multiple robots. The core principles of speed, security, and simplicity will guide our future improvements.
\vspace{-5mm}
\section{Related Work}
\subsection{Wearable BCIs to Assist People Daily Tasks}
\textcolor{black}{The HCI community has been interested in building user-friendly and effective brain-computer interaction mechanisms in recent years. With the rapid development of wearable devices and dry-electrode (i.e., non-intrusive) techniques, it is possible now to develop wearable EEG application devices to help people with motion disabilities. In terms of wearable devices, brain waves are utilized to form preliminary EEG signals. Brain waves can be broadly classified into five patterns: Delta (0.1-4 HZ), Theta (4-9 HZ), Alpha (8-12 HZ), Beta (12-30 HZ), and Gamma (above 30 HZ), corresponding to five mental modes~\cite{2009Rhythms}. Some good studies found out the potential of BCI applications using medical-grade devices, for instance, EEG-based wearable knee exoskeletons could rehabilitate the gait and restore the function for muscular disabilities related to knee motion, like standing up and sitting down~\cite{VILLAPARRA20151379}. Additionally, a language or hand-disabled person wearing such a device could detect his or her emotions and send them to the service robot~\cite{huang2014novel}. Nowadays, a civil EEG device is an essential option for wide use. Currently, companies like EPOC~\cite{Emotiv} and NeuroSky~\cite{brainlink} are designing advanced wireless and portable EEG systems in the form of wearable headphones. People with limited mobility can use these civil EEG devices with minimal technical supervision. However, due to the weak signal, low signal-to-noise ratio, low source accuracy, and low spatial resolution of civil EEG devices, it is difficult to achieve accurate intention detection, which may not meet the requirements of developing BCI solutions for accessible applications. These limitations might be circumvented with specific processing, such as extracting more EEG-related semantic information from EEG signals through an advanced network. }

\subsection{\textcolor{black}{Deep Learning Networks in BCI}} As the primary task of BCI is to encode brain thinking into command in a control system, the processing and recognition of EEG signals are also one of the most important fields in BCI research in recent years. By leveraging the stability of electrocorticography (ECoG) interfaces and neural plasticity, Silversmith et al. offered an approach for reliable, stable BCI control ~\cite{SakhaviTNNLS2018}. Recently, many deep learning-based methods have been applied and performed well in EEG recognition tasks~\cite{LIubicomp2022}. Tao et al. proposed an attention-based convolutional recurrent neural network (ACRNN) to extract more discriminative features from EEG signals and improve the accuracy of emotion recognition ~\cite{TaoTAC2020}. Sakhavi et al. proposed a classification framework for motor imagery data by introducing a new temporal representation of the data and also utilizing a CNN architecture for classification, where the new representation is generated from modifying the filter-bank common spatial patterns method~\cite{SakhaviTNNLS2018}. In the meantime, the high computing power and deployment requirements of deep learning method also bring challenges to its application in the BCI area. How to decrease the temporal and spatial consumption of deep learning models is significant for BCI as well. To enable each deep learning model to offer flexible resource-accuracy trade-offs, NestDNN as a framework takes the dynamics of runtime resources into account to enable resource-aware multi-tenant on-device deep learning for mobile vision systems~\cite{FangMobiCom2018}.

\subsection{EEG-based Cyber-Physical Control Systems}
\textcolor{black}{Diverse studies have explored using EEG for the control of various cyber-physical systems. Notably, Doud et al. devised an inventive system enabling users to seamlessly navigate a virtual helicopter using motor imagery-based BCI. Subsequently, they extended their approach by developing a noninvasive scalp EEG system for steering a robotic AR Drone quadcopter in physical space, all based on BCI2000~\cite{lafleur2013quadcopter}. Li and his team built an EEG-based BCI system using an Emotive EPOC neuroheadset, a laptop, and a Lego Mindstorms NXT robot \cite{li2016towards}, enabling participants to guide a simulated vehicle indoors through the utilization of steady-state visual evoked potentials (SSVEP) and motor imagery (MI) BCI systems. The accuracy of EEG signals has become a more significant concern When the task becomes more complicated. In a study employing the P300 BCI interface and SSVEP pattern to discern human intentions, the system's offline accuracy is a mere 55.56\%, and the recognizable target is limited to within the visual range of the robot~\cite{mao2019brain}. Another study~\cite{braun2019prototype} used BCI to support people and achieve an online detection accuracy of 70-90\%. The low accuracy is caused by low bandwidth for communication channels and EEG-related semantic loss. To improve accuracy, Chai et al. \cite{chai2013brain} presented a network for the three-class EEG task to control the wheelchair. To overcome the challenge of low bandwidth for the communication channel, Gandhi et al. proposed a user-centric adaptive interface based on an adaptive shared control mechanism \cite{gandhi2014eeg}.  
To overcome the shortcomings in interaction, more advanced methods are needed to be involved in the system, such as a deeper learning network of EEG data, a more advanced encoding-decoding communication, and heterogeneous characteristics engineering for the better capacity of feature extraction.}

\subsection{\textcolor{black}{Encoding-Decoding Communication Framework}} The semantic information is widely used in fields of NLP, data security, computer vision, etc., and semantic-based encoding-decoding communication is one of the emerging encoder-decoder communication paradigms that focuses on extracting semantic information from the transmitted message rather than receiving accurate bytes like traditional communication frameworks~\cite {QinArxiv2022}. Jiang \textit{et al.} presented XBee, a unique receiver-side Cross-technology Communication (CTC) achieving CTC bidirectionality, where the key innovation lies in the unique mechanism of cross-technology decoding~\cite{JiangMobiCom2018}. He \textit{et al.} proposed VI-Eye, the first communication framework that can align vehicle-infrastructure point clouds at centimeter accuracy in real-time, which could exploit traffic domain knowledge by detecting a set of key semantic objects including road, lane lines, curbs, and traffic signs~\cite{HeMobiCom2021}. Benefitting from the advances in AI technology, encoding-decoding communication research using deep learning has emerged last few years~\cite{LuoIWC2022}. Xie \textit{et al.} proposed a semantic-based encoding-decoding communication framework, named DeepSC, for text transmission. Based on the Transformer, the DeepSC aims at maximizing the framework capacity and minimizing the semantic errors by recovering the meaning of sentences~\cite{XieTSP2021}. Specifically for multi-user scenarios, Xie \textit{et al.} investigated multi-user encoding-decoding communication frameworks for transmitting single-modal semantic data and multimodal semantic data~\cite{XieJSAC2022}, respectively. Shi \textit{et al.} proposed a novel architecture based on federated edge intelligence for supporting resource-efficient semantic-aware networking~\cite{ShiComMag2021}.

\section{CONCLUSION}
In this paper, we introduced the design, implementation, evaluation of BRIEDGE, a system that enables multiple users to interact with multiple robots through EEG-adaptive neural networks and encoding-decoding communications. BRIEDGE takes the joint features extraction of heterogeneous EEG data into consideration and addresses the parallel transmissions under multi-user and multi-task scenarios to jointly maximize their performance. In addition, the BRIEDGE can be easily deployed on the edge AI  by adopting the model compression mechanism. Our evaluations show that BRIEDGE outperforms the other baselines in classification accuracy and signal transmission. We believe that BRIEDGE has taken a significant step towards turning the multi-brain to multi-robot interaction and collaboration tasks into reality.

\bibliographystyle{ACM-Reference-Format}
\bibliography{arxiv111}

\appendix

\end{document}